\documentclass[10pt,aps,physrev,reprint,floatfix,superscriptaddress,longbibliography]{revtex4-2}
\pdfoutput=1
\usepackage{graphicx,latexsym}
\usepackage{dcolumn,amsfonts}
\usepackage{amssymb,amsmath,bm}

\usepackage[breaklinks=true,pdfencoding=auto]{hyperref}

\usepackage{natbib}
\usepackage{balance}

\hypersetup{
	colorlinks   = true, %Colours links instead of ugly boxes
	urlcolor     = blue, %Colour for external hyperlinks
	linkcolor    = blue, %Colour of internal links
	citecolor    = red %Colour of citations
}

\usepackage{color}
\usepackage{ulem}
\begin{document}

%-----------------------------------------------------------------
	\title{The effects of a far-infrared photon cavity field on the  magnetization\break
		of a square quantum dot array}

	\author{Vidar Gudmundsson}
	\email{vidar@hi.is}
	\affiliation{Science Institute, University of Iceland, Dunhaga 3, IS-107 Reykjavik, Iceland}
	\author{Vram Mughnetsyan}
	\email{vram@ysu.am}
	\affiliation{Department of Solid State Physics, Yerevan State University, Alex Manoogian 1, 0025 Yerevan, Armenia}
	\author{Nzar Rauf Abdullah}
	\affiliation{Physics Department, College of Science,
		University of Sulaimani, Kurdistan Region, Iraq}
	\affiliation{Computer Engineering Department, College of Engineering, Komar University
		of Science and Technology, Sulaimani 46001, Kurdistan Region, Iraq}
	\author{Chi-Shung Tang}
	\email{cstang@nuu.edu.tw}
	\affiliation{Department of Mechanical Engineering, National United University, Miaoli 36003, Taiwan}
	\author{Valeriu Moldoveanu}
	\email{valim@infim.ro}
	\affiliation{National Institute of Materials Physics, PO Box MG-7, Bucharest-Magurele,
		Romania}
	\author{Andrei Manolescu}
	\email{manoles@ru.is}
	\affiliation{Department of Engineering, Reykjavik University, Menntavegur
		1, IS-102 Reykjavik, Iceland}

%
%----------------------------------------------------------------

\begin{abstract}
The orbital and spin magnetization of a cavity-embedded quantum dot array defined in a
GaAs heterostructure are calculated within quantum-electrodynamical density-functional theory (QEDFT).
To this end a gradient-based exchange-correlation functional recently employed for atomic systems is
adapted to the hosting two-dimensional electron gas (2DEG) submitted to an external perpendicular
homogeneous magnetic field. Numerical results reveal the polarizing effects of the cavity photon field
on the electron charge distribution and nontrivial changes of the orbital magnetization.
We discuss its intertwined dependence on the electron number in each dot, and on the electron-photon
coupling strength. In particular, the calculated dispersion of the photon-dressed electron states
around the Fermi energy as a function of the electron-photon coupling strength indicates the formation
of magnetoplasmon-polaritons in the dots.
\end{abstract}

\maketitle
%
%----------------------------------------------------------------------------------------
%

\section{Introduction}
Research into the influence of cavity photon modes on electron transport
through nanoscale systems \cite{Paravicini-Bagliani2019},
optical properties of two dimensional electron systems \cite{Yoshie2004,Zhang1005:2016,PhysRevX.7.011030},
quasi-particle excitation in light-matter systems \cite{PhysRevB.49.8774,PhysRevLett.90.116401,Ciuti05:115303,PhysRevLett.111.176401},
or processes in chemistry \cite{https://doi.org/10.1002/anie.201107033,doi:10.1126/science.abd0336,doi:10.1126/science.aau7742,Schafer2021ShiningLO},
has been gaining attention in the last three decades, just to cite the work of few groups involved.
The diverse systems and their phenomena have been theoretically described by a multitude of methods
ranging from ``simple toy models'' \cite{Jaynes63:89,Bishop2009},
nonequilibrium Green functions \cite{PhysRevA.87.023831}, and master equations of various types.
In the more complex models with few to many charged entities, traditional approaches to many-body theory, or Configuration Interactions (CI) (exact diagonalization in many-body Fock space) have been used
\cite{Entropy19:731}, but relatively recently Density Functional Theory approaches have been appearing
\cite{PhysRevA.90.012508,PhysRevA.98.043801,FlickRiveraNarang+2018+1479+1501,doi:10.1021/acsphotonics.9b00768}.

{The foundation of the approach lies in combining the polarization of a Dirac or a
Schr{\"o}dinger field and the vector potential for a photon field by solving two coupled nonlinear differential equations for the evolution without
explicitly referring to many-body wavefunctions of the coupled system \cite{PhysRevA.90.012508}.
The procedure has been illustrated excellently by Flick et al in a more recent publication \cite{FlickRiveraNarang+2018+1479+1501}.}

Only very recently, an explicit analytical gradient-based functional exchange-correlation energy
functional derived from the adiabatic-connection fluctuation-dissi\-pation theorem
\cite{PhysRevA.81.062708} has been
published \cite{flick2021simple}. Interestingly, the development of this functional is closely
related to earlier work on the van der Waals and the Casimir interactions in complex
molecular systems and macroscopic bodies \cite{PhysRevLett.118.266802,PhysRevLett.103.063004}.

Experiments have shown that a two-dimensional electron gas (2DEG) in a GaAs heterostructure
is an ideal experimental system to achieve a strong electron-photon coupling in a
FIR cavity \cite{Zhang1005:2016}.
Motivated by this fact we suitably modify the exchange and correlation energy functional
and extend it to a square periodic lattice of quantum dots formed in a two-dimensional
electron gas (2DEG) and subjected to perpendicular homogeneous magnetic field.
We apply it to investigate the influences of the
coupling to a cavity photon mode on the orbital magnetization of the system. This choice of
applications has two intertwined reasons: First, the orbital and the spin components of the magnetization
are equilibrium quantities, that do not invoke any need of calculations of the dynamical properties
of the system \cite{Meinel99:819,Meinel01:121306}.
Second, the magnetization is thus an appropriate experimental equilibrium quantity,
that could be used to control and assist in the development of Quantum-Electrodynamical Density
Functional Theory (QEDFT) approaches, that are needed for wide areas of applications in
physics and chemistry.

In addition, it is clear that the polarization of the electron density
by the cavity-photon field must change the orbital magnetization of the electron system,
and as GaAs systems are ideal to enable tuning of the electron-photon coupling strength we
do the calculations for GaAs parameters \cite{Zhang1005:2016}.

In the Model Section \ref{Model} we introduce the computational model for the 2DEG
collecting all technical, but important, details on the Local Spin Density Field Theoretical
(LSDA) approach for the electron-electron Coulomb interaction to Appendix \ref{e-Coulomb}. Also,
details about the construction of the gradient-based nonlocal exchange and correlation functional
for the electron-photon interaction are given in Appendix \ref{e-EM-functionals}. The results are
presented in the Section \ref{Results}, and conclusions are summarized in
Section \ref{Conclusions}.

\section{Model}
\label{Model}
We model electrons in two-dimensional (2D) square periodic lattice of quantum
dots defined by the potential
\begin{equation}
       V_\mathrm{per}(\bm{r}) = -V_0\left[\sin \left(\frac{g_1x}{2} \right)
	   \sin\left(\frac{g_2y}{2}\right) \right]^2,
	   \label{Vper}
\end{equation}
with $V_0 = 16.0$ meV. The dot square array is defined by the lattice vectors
$\bm{R}=n\bm{l}_1+m\bm{l}_2$, where $n,m\in \bm{Z}$. Its unit vectors
are $\bm{l}_1 = L\bm{e}_x$ and $\bm{l}_2 = L\bm{e}_y$.
The reciprocal lattice is spanned by $\bm{G} = G_1\bm{g}_1 + G_2\bm{g}_2$ with
$G_1, G_2\in \bm{Z}$ with the unit vectors
\begin{equation}
	\bm{g}_1 = \frac{2\pi\bm{e}_x}{L}, \quad\mbox{and}\quad
	\bm{g}_2 = \frac{2\pi\bm{e}_y}{L},
\end{equation}
where $L = 100$ nm. A local spin-density functional theory (LSDFT) approach is used to
describe the mutual Coulomb interactions of the electrons in the presence of a homogeneous
external magnetic field $\bm{B}=B\hat{\bm e}_z$ perpendicular to the plane of the two-dimensional electrons gas (2DEG). In order to fulfill the commensurability conditions of the lattice length
of the dot lattice and the characteristic length scale associated with the magnetic
field $l=(\hbar c/(eB))^{1/2}$ \cite{Harper55:874,Azbel64:634,Hofstadter76:2239}
we use a state basis constructed by
Ferrari \cite{Ferrari90:4598} in the symmetric gauge, and used previously by Silberbauer
\cite{Silberbauer92:7355} and Gudmundsson \cite{Gudmundsson95:16744}.
The QEDFT Hamiltonian of the Coulomb interacting electrons in a photon cavity is
\begin{equation}
	H = H_0 + H_\mathrm{Zee} + V_\mathrm{H} + V_\mathrm{per} + V_\mathrm{xc} + V^\mathrm{EM}_\mathrm{xc},
	\label{H}
\end{equation}
where the last term describing the coupling of the electrons to the cavity photon field
will be described in details later. $H_0$ is the Hamiltonian of free 2D electrons in the
external magnetic field
\begin{equation}
 	H_0 = \frac{1}{2m^*}\bm{\pi}^2, \quad\mbox{with}\quad
 	\bm{\pi} = \left(\bm{p}+\frac{e}{c}\bm{A} \right).
 	\label{H0}
\end{equation}
In the symmetric gauge the vector potential is $\bm{A}= (B/2)(-y,x)$.
The wavefunctions of the Ferrari basis \cite{Ferrari90:4598} are the Kohn-Sham
eigenfunctions of $H_0$ denoted by $\phi^{\mu\nu}_{n_l}(\bm{r})$ with $n_l=0,1,2,\cdots$ a
Landau band index, and  $\mu = (\theta_1+2\pi n_1)/p$, $\nu = (\theta_2+2\pi n_2)/q$,
with $n_1\in I_1 = \{0,\dots, p-1 \}$, $n_2\in I_2 = \{0,\dots, q-1 \}$,
and $\theta_i\in [-\pi,\pi]$. $pq$ is the number of magnetic flux quanta $\Phi_0 = hc/e$
flowing through a unit cell of the lattice. We denote the eigenfunctions of the total
Hamiltonian with $\psi_{\bm{\beta\theta}\sigma}(\bm{r})$, where $\sigma =\{\uparrow\downarrow\}$
indicates the $z$-component of the spin, $\bm{\theta}=(\theta_1,\theta_2)$ is the location
in the unit cell of the reciprocal lattice, and stands for all remaining quantum numbers.
In each point of the reciprocal lattice, $\bm{\theta}$, the eigenfunctions $\phi^{\mu\nu}_{n_l}$
and $\psi_{\bm{\beta\theta}\sigma}$ form complete orthonormal bases if
$(\mu,\nu )\neq (\pi,\pi )$ for all $(n_1,n_2)\in I_1\times I_2$.

The Hartree part of the Coulomb interaction is
\begin{equation}
	V_\mathrm{H}(\bm{r}) = \frac{e^2}{\kappa}\int_{\bm{R}^2}d\bm{r}'\frac{\Delta n(\bm{r}')}
	{|\bm{r}-\bm{r}'|},
	\label{Vcoul}
\end{equation}
with $\Delta n(\bm{r}) = n(\bm{r})-n_\mathrm{b}$, where $+en_\mathrm{b}$ is the
homogeneous positive background charge density needed to maintain the total system charge neutral.
We consider the positive background charge to be located in the plane of the 2DEG.
The electron density is
\begin{align}
	  n_e(\bm{r}) =& n_\uparrow(\bm{r}) + n_\downarrow(\bm{r})\nonumber\\
	  =& \frac{1}{(2\pi)^2}\sum_{\bm{\alpha}\sigma}
	  \int^{\pi}_{-\pi}d\bm{\theta}\; \left|\psi_{\bm{\beta\theta}\sigma}(\bm{r}) \right|^2
	  f(E_{\bm{\beta\theta}\sigma}-\mu),
\label{ne}
\end{align}
where the $\bm{\theta}$-integration is over the two-dimensional unit cell in
reciprocal space, $f$ is the Fermi equilibrium distribution with the chemical potential $\mu$
and $E_{\bm{\beta\theta}\sigma}$ the energy spectrum of the QEDFT Hamiltonian $H$ (\ref{H}), where $\sigma = \uparrow$ or
$\downarrow$ is the spin label.
We choose the temperature $T=1.0$ K.
The Zeeman Hamiltonian is $H_\mathrm{Zee}=\pm g^*\mu^*_\mathrm{B}B/2$, and we use GaAs parameters
$m^*=0.067m_e$, $\kappa = 12.4$, and $g^*=0.44$.

The potential describing the exchange and correlation effects of the Coulomb interaction
of the electrons
\begin{equation}
	V_{\mathrm{xc},\sigma}(\bm{r},B)=\frac {\partial}{\partial n_{\sigma}}
	(n_e\epsilon_\mathrm{xc}[n_{\uparrow},n_{\downarrow}, B])
	|_{n_{\sigma}=n_{\sigma}(r)},
	\label{V_xc}
\end{equation}
is derived from the Coulomb exchange and correlation functionals listed in Appendix \ref{e-Coulomb}.
The exchange and correlation contribution to the Coulomb interaction and the positive background
charge make the localization of several electrons possible in a quantum dot.

In Appendix \ref{e-EM-functionals} we detail how we adopt the exchange and correlation
functional for a 2DEG in a perpendicular magnetic field and modify slightly the gradient-based
exchange and correlation functional presented recently by Flick \cite{flick2021simple} for
the interaction of electrons with the photon field in a cavity. We assume the reflective
plates of the cavity to be parallel to the 2DEG and use the dipole approximation assuming
the wavelength of the FIR-field to be much lager than the lattice length $L$.
{We furthermore assume the cavity plates to be disks in order not to promote
a preferred polarization of the electric component of the cavity photon field. We emphasize
that even though we do select one photon mode here nothing is specified about the number of
photons present in the cavity in different situations.}
The functional can be expressed as
\begin{align}
	E_\mathrm{xc}^\mathrm{GA}[n,&\bm{\nabla}n] = \frac{1}{16\pi}\sum^{N_p}_{\alpha = 1}
	|\lambda_\alpha|^2 \nonumber\\
	&\int d\bm{r} \frac{\hbar\omega_p(\bm{r})}{\sqrt{(\hbar\omega_p(\bm{r}))^2/3+
	(\hbar\omega_g(\bm{r}))^2}+\hbar\omega_\alpha},
\label{E_EM_xc}
\end{align}
where $\hbar\omega_\alpha$ and $\lambda_\alpha$ are the energy and coupling strength
of cavity-photon mode $\alpha$, respectively. $N_p$ is the number of cavity modes.
The coupling strength is expressed here in
units of meV$^{1/2}$/nm as is explained in Appendix \ref{e-EM-functionals}.
The gap-energy \cite{flick2021simple,PhysRevLett.103.063004,PhysRevA.81.062708} stemming
from considerations of dynamic dipole polarizability leading to the van der Waals interaction
is given by
\begin{equation}
	\left(\hbar\omega_g(\bm{r})\right)^2 = C \left|\frac{\bm{\nabla}n_e}{n_e}\right|^4\frac{\hbar^2}{{m^*}^2}
	= C(\hbar\omega_c)^2l^4\left|\frac{\bm{\nabla}n_e}{n_e}\right|^4
\label{hwgr}
\end{equation}
with $C=0.0089$ and the cyclotron frequency $\omega_c=eB/(m^*c)$.
We remind the reader here, that even though the electron gas is
considered two-dimensional, the electrodynamics remain three-dimensional.
The dispersion of a magnetoplasmon confined to two dimensions is
\cite{PhysRevLett.18.546,Ando82:437}
\begin{equation}
	\left(\hbar\omega_p(q)\right)^2 = (\hbar\omega_c)^2 + \frac{2\pi n_e^2}{m^*\kappa}q + \frac{3}{4}v_\mathrm{F}^2q^2,
	\label{hwpq}
\end{equation}
where $v_\mathrm{F}$ is the Fermi velocity, and $q$ is a general wavevector. Importantly, the
plasmon at low magnetic field is almost gap-less indicating that the 2DEG is ``softer'', regarding
external perturbation, than a three-dimensional electron gas. To construct a local plasmon dispersion
we use the commonly used relation in this context that
$q\approx k_\mathrm{F}/6\approx |\bm{\nabla}n_e|/n_e$ \cite{PhysRevB.33.8800,PhysRevLett.47.446}
and obtain
\begin{align}
	\left(\hbar\omega_p(\bm{r})\right)^2 = &
	(\hbar\omega_c)\left(2\pi l^2n_e(\bm{r})\right)\left(\frac{e^2}{\kappa l}\right)
    \left(l\frac{|\bm{\nabla}n_e|}{n_e} \right)\nonumber\\
    + & (\hbar\omega_c)^2\left\{ \frac{1}{36}\left(\frac{|l\bm{\nabla}n_e|}{n_e}\right)^4 + 1\right\},
\label{hwpr}
\end{align}
where the dimensional information is made explicit by collecting terms into bracket pairs.
Here, it becomes clear that for the 2D case we need all the terms specified for the magnetoplasmon
in order to keep the treatment of the gap-energy and the magnetoplasmon on equal footing.
The exchange and correlation potentials for the electron-photon interaction are generated using
the variation
\begin{equation}
	V_\mathrm{xc}^\mathrm{EM} = \frac{\delta E_\mathrm{xc}^\mathrm{GA}}{\delta n_e} =
	\left\{ \frac{\partial}{\partial n_e} -
	                  \bm{\nabla}\cdot\frac{\partial}{\partial\bm{\nabla}n_e}\right\}
	                  E_\mathrm{xc}^\mathrm{GA},
\label{V_EM_xc}
\end{equation}
together with the general extension to spin densities \cite{PhysRevB.33.8800}
\begin{equation}
    E_\mathrm{xc}[n_\uparrow,n_\downarrow] = \frac{1}{2}E_\mathrm{xc}[2n_\uparrow]
                                           + \frac{1}{2}E_\mathrm{xc}[2n_\downarrow]
\label{spin-d-d}
\end{equation}
and
\begin{equation}
	\frac{\delta E_\mathrm{xc}[n_\uparrow,n_\downarrow]}{\delta n_\sigma} =
	\left.\frac{\delta E_\mathrm{xc}[n_e]}{\delta n_e(\bm{r})}\right|_{n_e(\bm{r})=2n_\sigma(\bm{r})}.
\label{delta-2}
\end{equation}
{The electron-photon exchange-correlation potentials (\ref{V_EM_xc}) are added to the
DFT self-consistency iterations.}

As the equilibrium electron spin densities are periodic all matrix elements can be related to
analytical matrix elements of phase factors $\exp{(-i\bm{G}\cdot\bm{r})}$ in the original Ferrari basis
\cite{Ferrari90:4598,Gudmundsson95:16744}. Thus it is convenient to construct the various
gradient terms of the electron spin densities using their Fourier transforms in order to minimize
the computational errors.

{In constructing the exchange and correlation functional (\ref{E_EM_xc}), Flick combines
contributions due to the para- and the diamagnetic electron-photon interactions (\ref{QEDFT-Hint})
up to second order in the coupling strength, noting that ``physically such an approximation
corresponds to including one-photon exchange processes
explicitly, while neglecting higher order processes'', to use his words \cite{flick2021simple}.
Importantly, the self-consistency requirement in our calculations thus bring into play higher order
effects of repeated single photon processes.}

The orbital and the spin components of the magnetization of the system are
\begin{align}
	M_o+M_s=&\frac{1}{2c{\cal A}}\int_{\cal A} d\bm{r}\: \left( {\bf r}\times
	 {\bf j}({\bf r}) \right) \cdot\hat{{\bf e}}_z
	\nonumber\\
	-&\frac{g^*\mu_B^*}{{\cal A}}\int_{\cal A} d\bm{r}\:  \sigma_z
	({\bf r}),
	\label{M_OS}
\end{align}
where ${\cal A}=L^2$ is the area of a unit cell in the system, and the current density is given by
Eq.\ (\ref{currD}) in Appendix \ref{e-EM-functionals}. The magnetization is an equilibrium
quantity well suited to investigate the effects of the coupling of the FIR photon modes of the
cavity on.
%

%------------------------
\section{Results}
\label{Results}
For all calculations, unless use of other values is indicated, we use 10 Landau
bands (each split into $pq$ subbands), and a 16$\times$16 $(\theta_1,\theta_2)$-mesh for
a repeated 4-point Gaussian quadrature in the primitive zone of the reciprocal lattice.
For the spatial coordinates we use a 40$\times$40 $(x,y)$-mesh for a repeated 4-point
Gaussian quadrature. For all the calculations we consider only one FIR photon cavity mode with
energy $\hbar\omega_\alpha=1.0$ meV. For clarity we keep the notation for the energy of the mode as
$\hbar\omega_\alpha$ and the coupling as $\lambda_\alpha$, even though the index $\alpha$
could have been dropped.

\begin{figure}[htb]
	\centerline{\includegraphics[width=0.48\textwidth,bb=0 50 400 270]{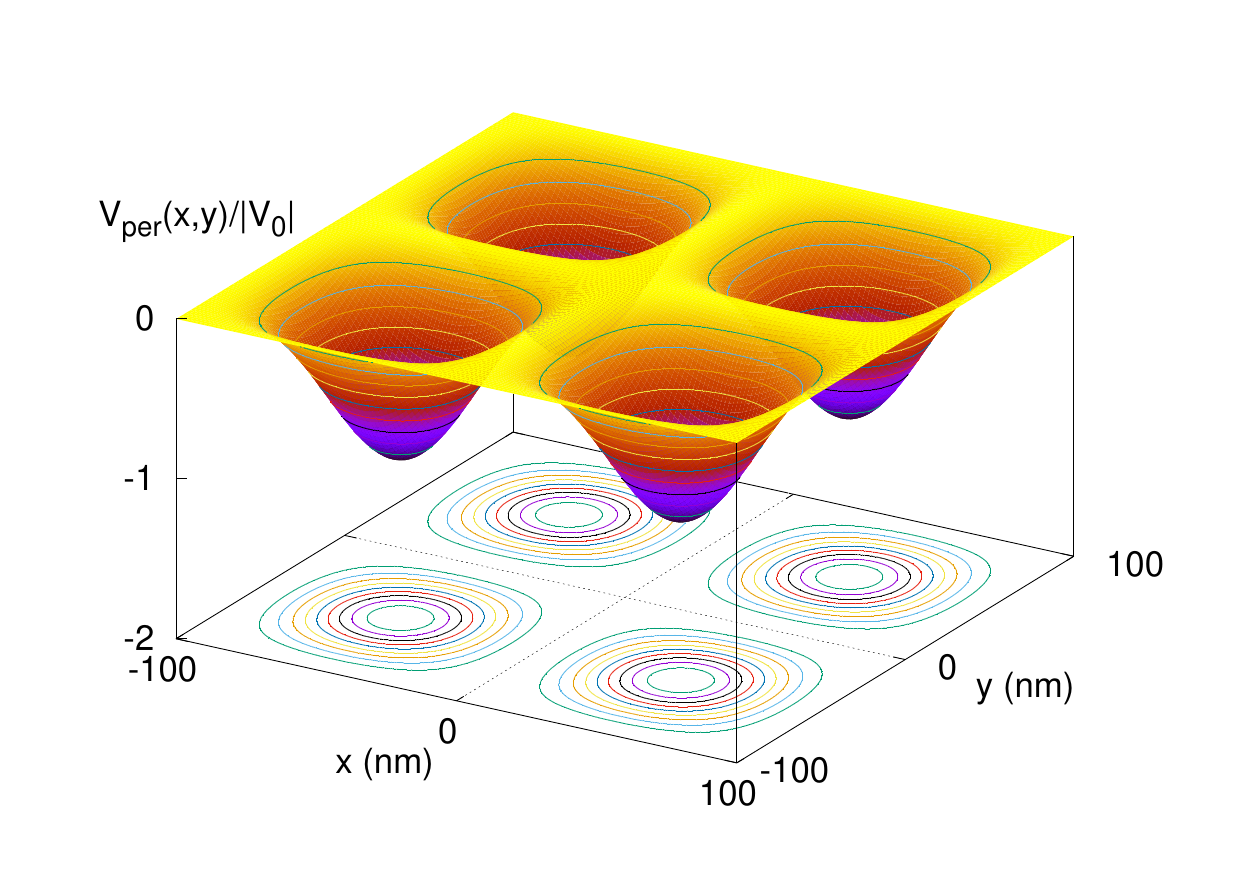}}
	\centerline{\includegraphics[width=0.48\textwidth,bb=0 0 400 220]{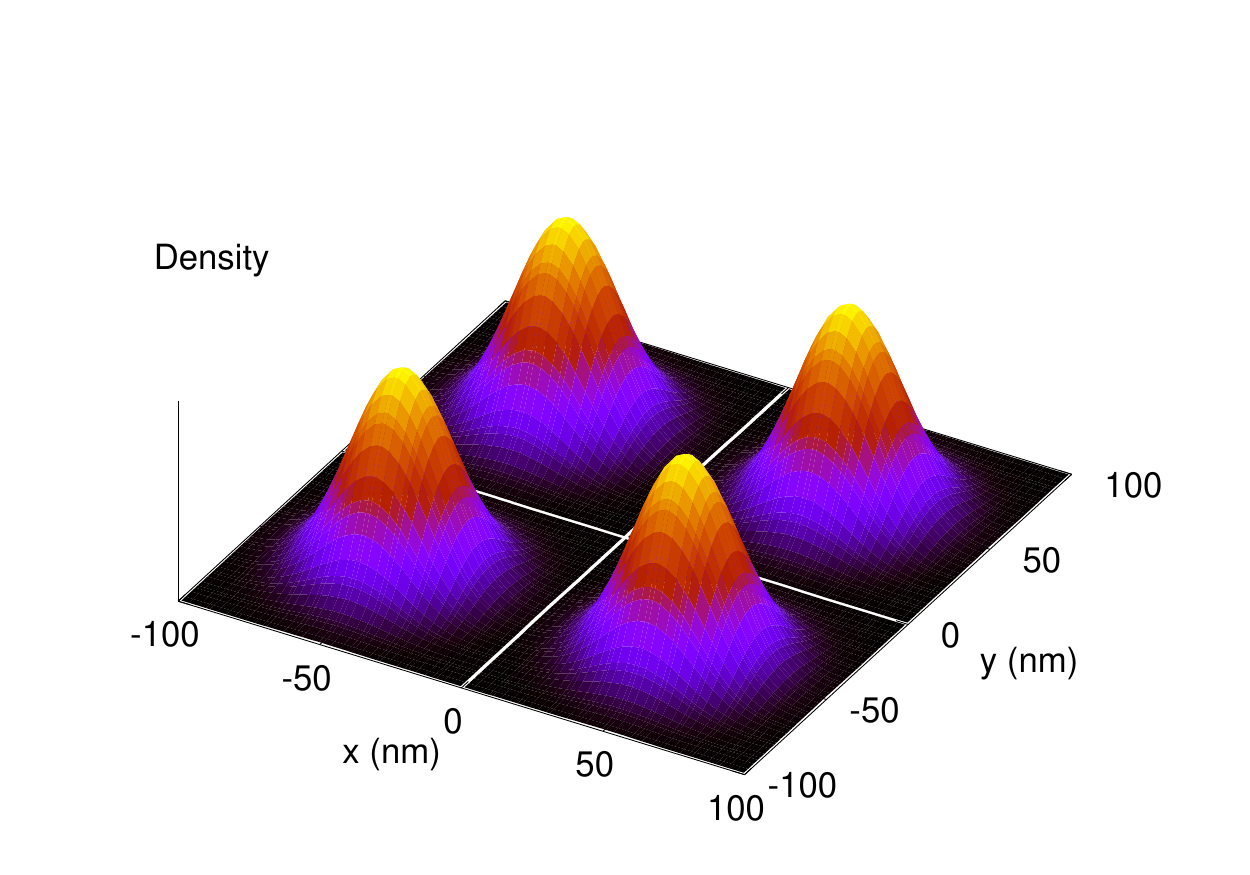}}
	\caption{For 4 unit cells of the square lattice, the dot confinement potential (upper panel)
		and the electron density for two electrons (lower panel), $N_e=2$, in each dot for $pq=1$ corresponding to
		$B=0.414$ T. $\lambda_\alpha l=0.050$ meV$^{1/2}$, $\hbar\omega_\alpha =1.0$ meV,
		$L=100$ nm, $\kappa = 12.4$, $T=1.0$ K,
		$V_0=-16.0$ meV, $g^*=0.44$ and $m^*=0.067m_e$.}
	\label{VxyDens}
\end{figure}

In the upper panel of Fig.\ \ref{VxyDens} the periodic dot potential is displayed for
4 unit cells of the square lattice. In the lower panel of the figure is the corresponding
electron density for two electrons, $N_e=2$, in each unit cell or dot for one quantum
of magnetic flux, $pq=1$, through a unit cell corresponding to the magnetic field $B=0.414$ T.
The electron-photon coupling strength is $\lambda_\alpha l=0.050$ meV$^{1/2}$. Clearly, even at this low
magnetic field the overlap of the electron density between the dots is very low, i.e.\ the electrons
are well localized in the dots. This is facilitated by the homogeneous positive background charge
density in the plane of the 2DEG, and the effects of the exchange and correlation functional for
the electron-electron Coulomb interaction.

\begin{figure}[htb]
	\centerline{\includegraphics[width=0.48\textwidth,bb=0 50 370 265]{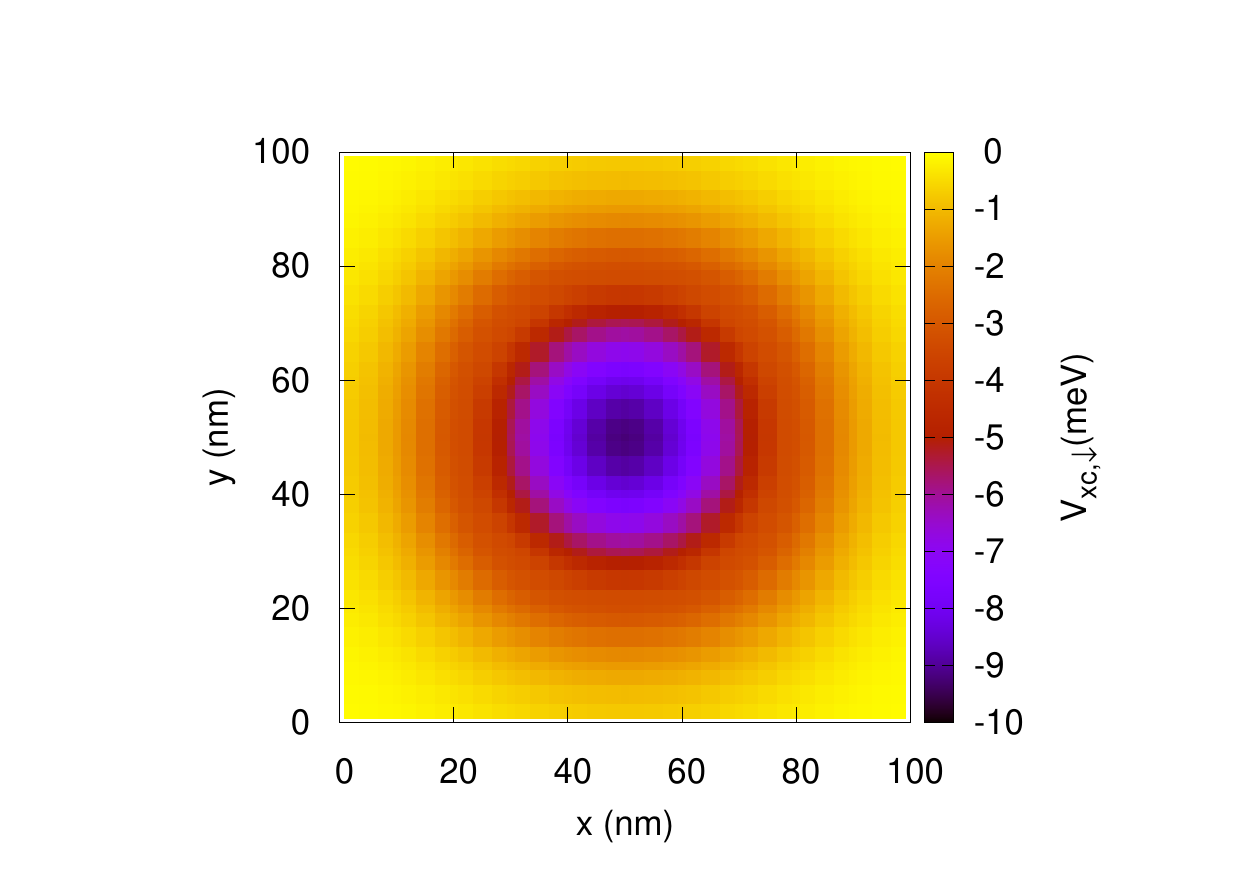}}
	\centerline{\includegraphics[width=0.48\textwidth,bb=0 10 370 265]{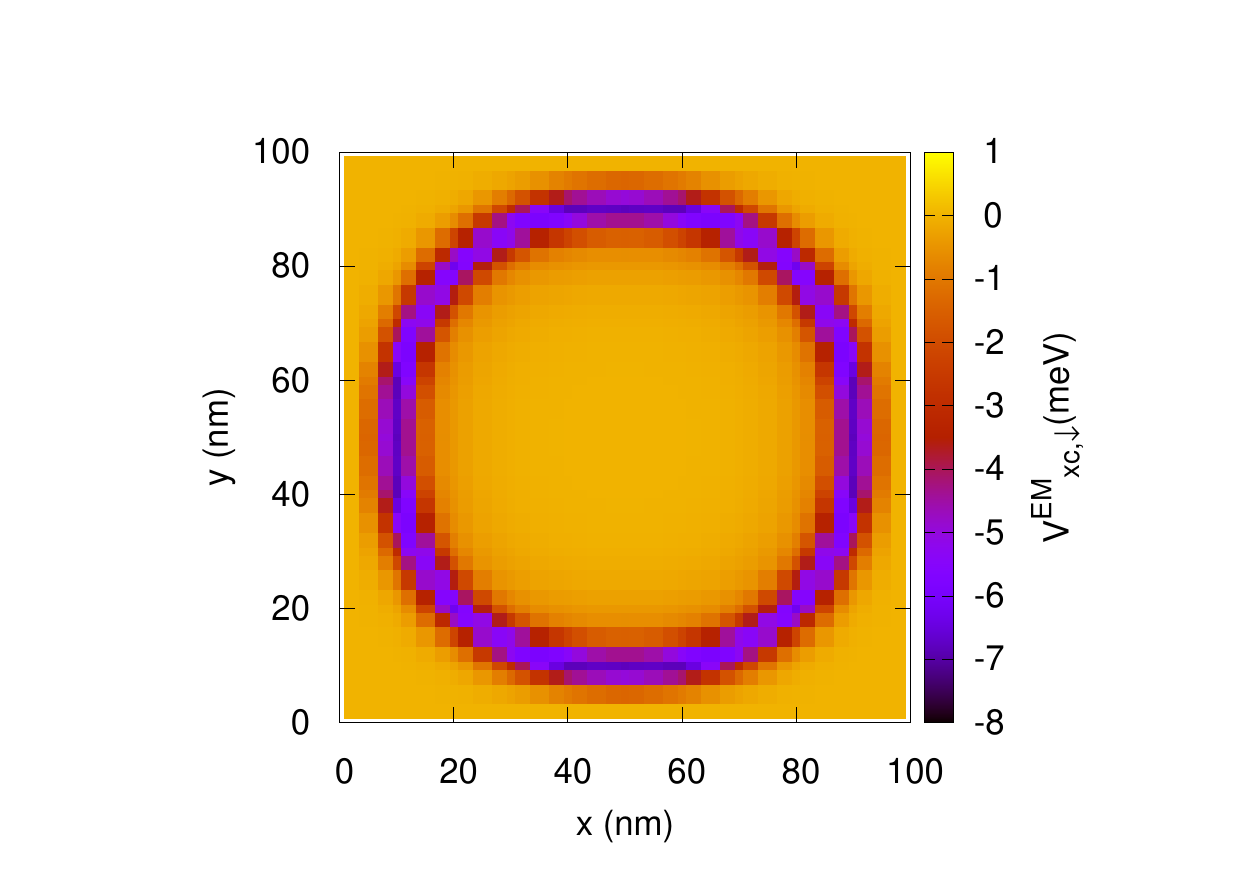}}
	\caption{The exchange and correlation potentials for the mutual electron Coulomb interaction
	         $V_\mathrm{xc}$ (upper panel), and the electron-photon interaction
	         $V^\mathrm{EM}_\mathrm{xc}$ (lower panel) for $N_e=2$, $pq=4$, $\lambda_\alpha l=0.050$ meV$^{1/2}$ and $\hbar\omega_\alpha =1.0$ meV. The potentials are shown for the lower energy
	         spin direction.}
	\label{VxcEM}
\end{figure}

For a higher magnetic field the two electrons are even better confined within the dot potential
as the magnetic length $l$ is then smaller compared to the lattice length $L$.
In the upper panel of Fig.\ \ref{VxcEM} the exchange and correlation potential
$V_\mathrm{ex,\downarrow}(\bm{r})$ (\ref{Vxc-Coul}) is shown for $pq=4$ and
$\lambda_\alpha l=0.050$ meV$^{1/2}$. Note the depth of the potential compared to the
maximum depth of the confinement potential, $V_0=-16$ meV.
In the lower panel of Fig.\ \ref{VxcEM} is the exchange and correlation potential
$V_\mathrm{xc,\downarrow}^\mathrm{EM}(\bm{r})$ (\ref{V_EM_xc}). The interaction with the
photon field with an isotropic distribution of a polarization vector tends to spread
the electron charge density as has been seen in calculations, where exact diagonalization
(or configuration interaction (CI)) methods have been used to describe the electron
cavity-photon interactions \cite{ANDP:ANDP201500298,Flick2017}. This action of
$V_\mathrm{xc,\downarrow}^\mathrm{EM}(\bm{r})$ to polarize or spread the charge explains
the attractive part seen in the lower panel of Fig.\ \ref{VxcEM} forming a ring structure
where the ratio of the gradients of the electron density versus the density itself are high.
This behavior is expected as such terms are numerous in the potentials. The relative sharpness
of the structures in $V_\mathrm{xc,\downarrow}^\mathrm{EM}(\bm{r})$ makes a strong requirement
of including a high number of Fourier coefficients in the numerical calculations, much higher
that is needed for $V_\mathrm{ex,\downarrow}(\bm{r})$. When the number of electrons in the
unit cell, $N_e$, is increased the potentials start to acquire square symmetry aspects from
the lattice with their characteristics deviating from the circular symmetry that has the upper
hand in Fig.\ \ref{VxcEM}.

\begin{figure}[htb]
	\centerline{\includegraphics[width=0.48\textwidth,bb=0 50 370 265]{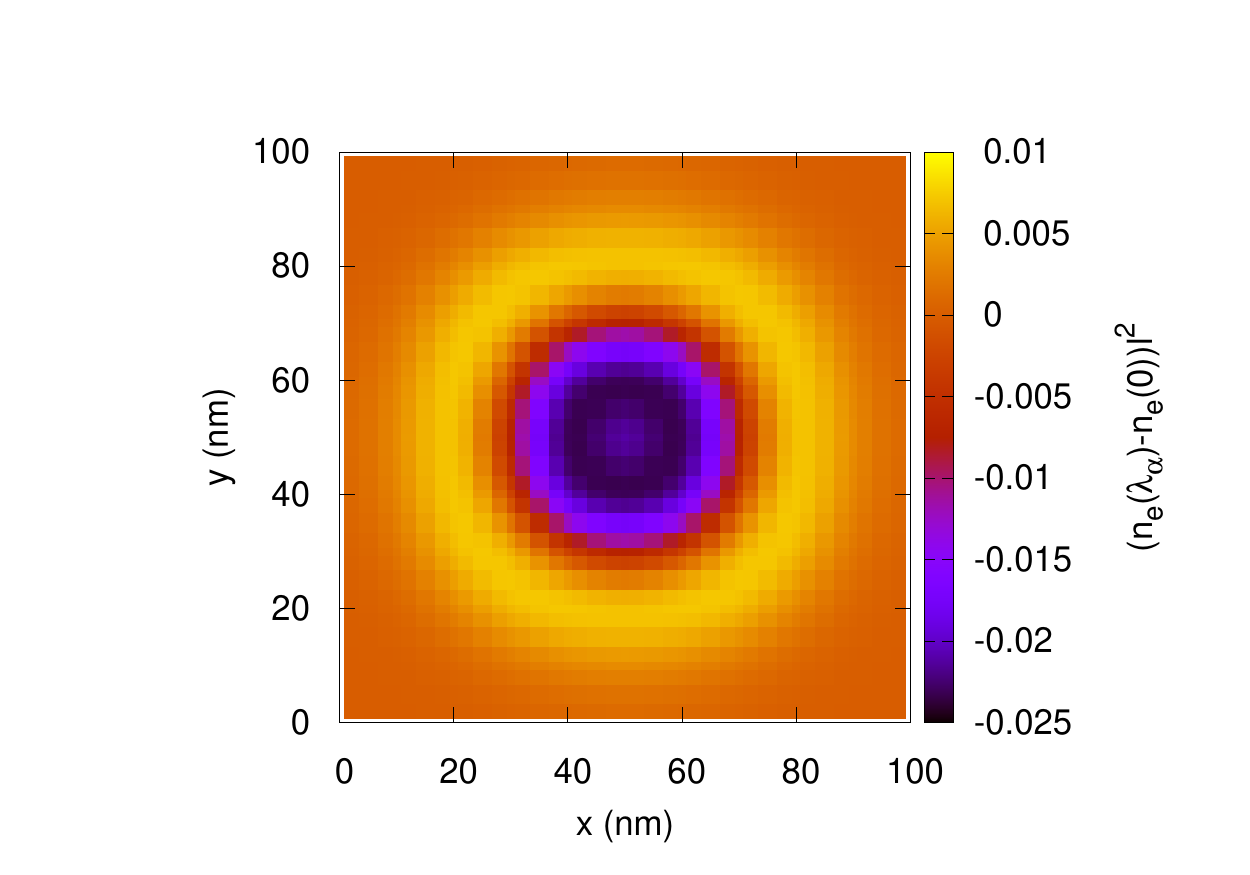}}
	\centerline{\includegraphics[width=0.48\textwidth,bb=0 10 370 265]{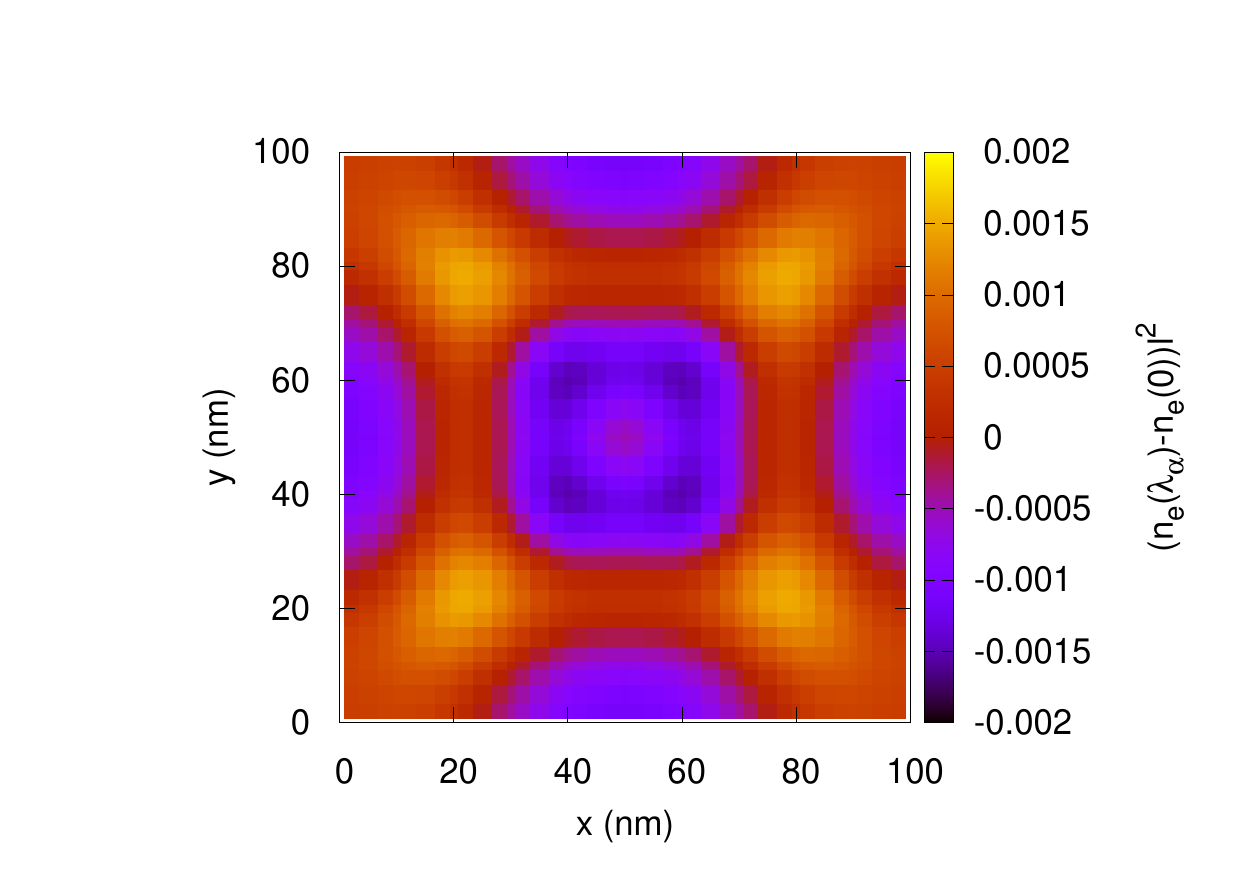}}
	\caption{The change in the electron density $[n_\mathrm{e}(\lambda_\alpha )-n_\mathrm{e}(0)]l^2$
	         for the electron-photon coupling $\lambda_\alpha l = 0.050$ meV$^{1/2}$
	         compared to the density without the coupling ($\lambda_\alpha l = 0$)
             for $N_\mathrm{e}=2$ (upper panel), and $N_\mathrm{e}=4$ (lower panel).
             $pq=4$, and $\hbar\omega_\alpha =1.0$ meV.}
	\label{ddn}
\end{figure}

In Fig.\ \ref{ddn} the polarization of the charge distribution defined as
$[n_\mathrm{e}(\lambda )-n_\mathrm{e}(0)]l^2$ is displayed for both $N_e=2$ (upper panel)
and $N_e=4$ (lower panel), with $\lambda_\alpha l = 0$ to 0.050 meV$^{1/2}$ and $pq=4$.
Clearly, for $N_e=2$, the polarization of the charge is isotropic like the distribution of the polarization vector of the photon field. Moreover, we see clearly how charge is moved away from
the center of the dot to its outskirts. The lower panel of Fig.\ \ref{ddn} shows that for the
higher number of electrons $N_e=4$ in the dots the charge polarized by the cavity photon field is not
anymore isotropic, but has assumed to a large extent the square symmetry of the underlying
square dot lattice. This is understandable as is is energetically favorable to relocate some
of the charge to the corner regions of the unit cell due to the strong direct Coulomb repulsion,
but still charge is polarized away from the center region of the dots.

\begin{figure}[htb]
	\includegraphics[width=0.32\textwidth,bb=0 60 208 326]{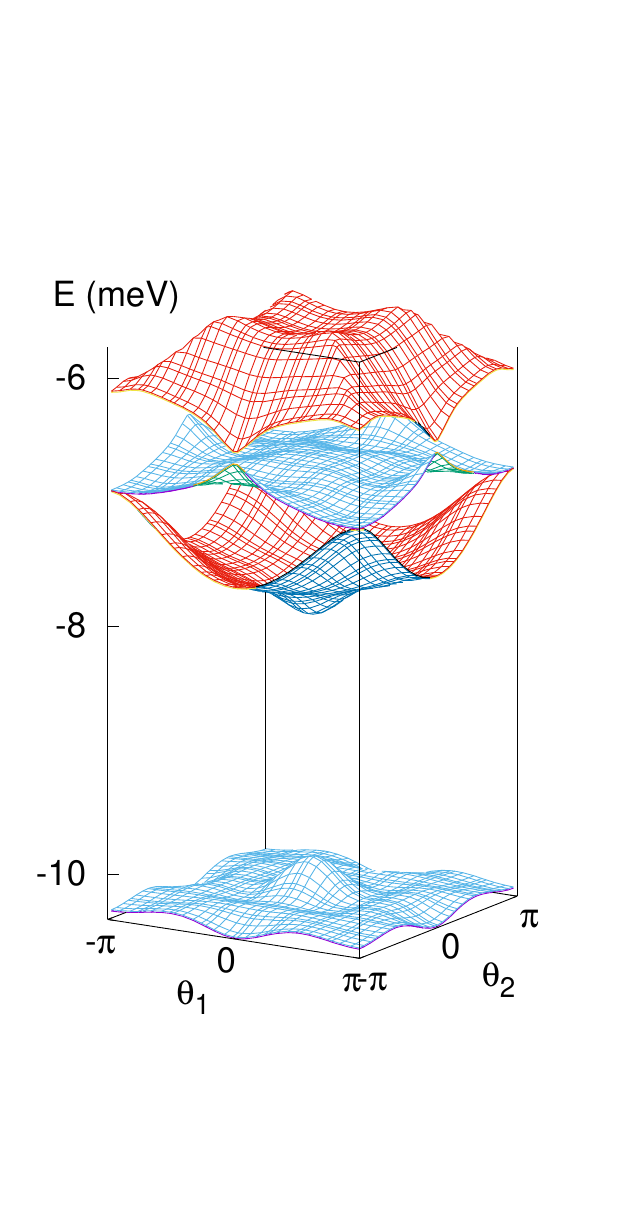}
	\caption{The 4 lowest in energy bands for $pq=1$ and $\lambda_\alpha l = 0.050$ meV$^{1/2}$.
	         Each band is composed of two spin subbands separated by the Zeeman energy
             $1.053\times 10^{-2}$ meV. As $N_\mathrm{e}=2$ the chemical potential
             $\mu = -8.954$ meV at $T=1.0$ K. $\hbar\omega_\alpha =1.0$ meV.}
	\label{Eband}
\end{figure}

We are investigating a periodic 2DEG here, so specially at high magnetic fields, we
expect the energy bands formed by the lattice periodicity to be narrow for the states
of electrons that are well localized in the dots, and higher up in the energy spectra
they can be expected to broaden. Exactly, this behavior can help us to get insight into
what is happening in the system as the electron photon coupling is changed.
To prepare for this analysis we show the 8 lowest energy bands for $N_e=2$ at $pq=1$
and $\lambda_\alpha l = 0.050$ meV$^{1/2}$ in Fig.\ \ref{Eband}.
Notice that for $pq=1$ $B=0.414$ T and for $g^*=0.44$ the spin splitting of the bands
is not clearly resolved in Fig.\ \ref{Eband}, but we see a large energy gap from the
2 lowest occupied bands to the higher unoccupied bands. The 2 lowest bands have only a
small dispersion, but the higher ones have a large one.

\begin{figure*}[htb]
	\includegraphics[width=0.48\textwidth,bb=0 50 400 300]{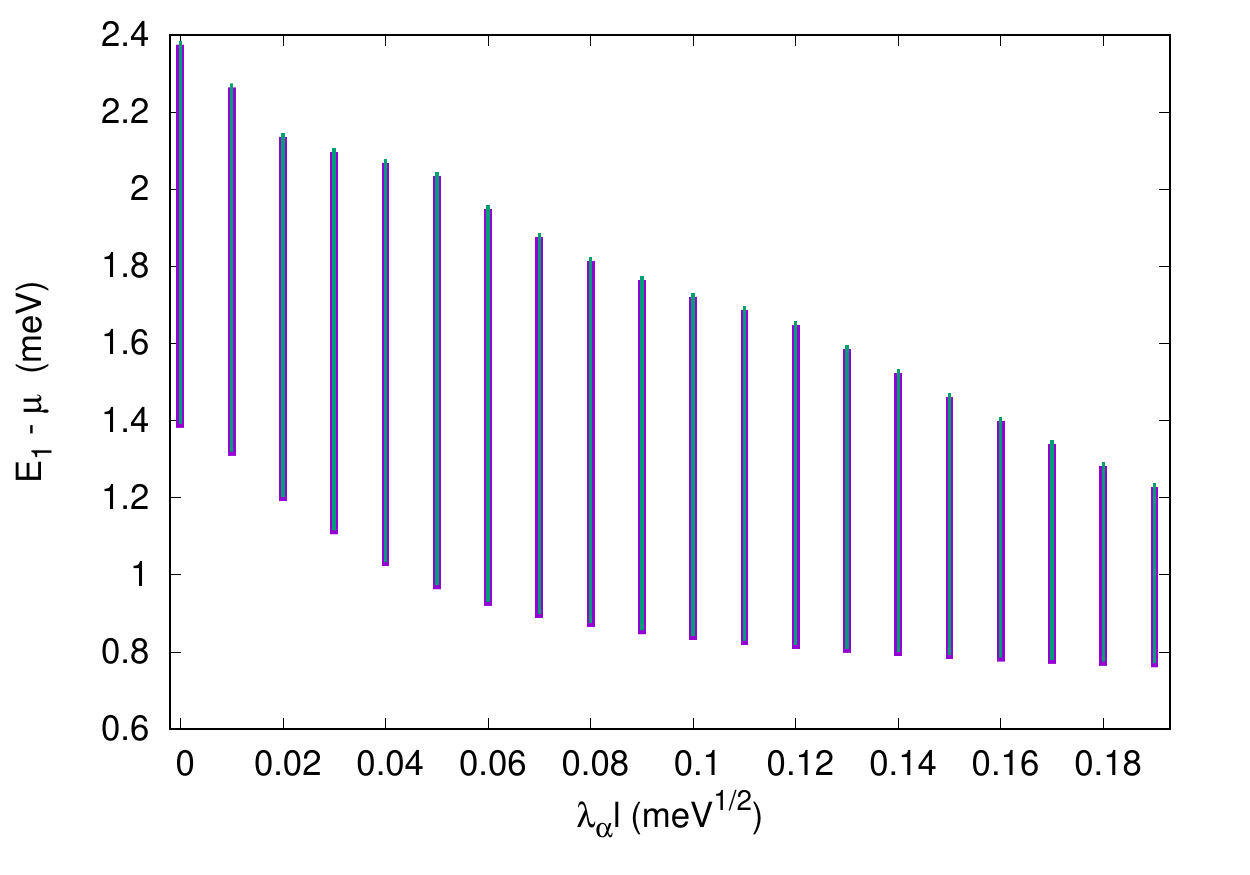}
	\includegraphics[width=0.48\textwidth,bb=0 50 400 300]{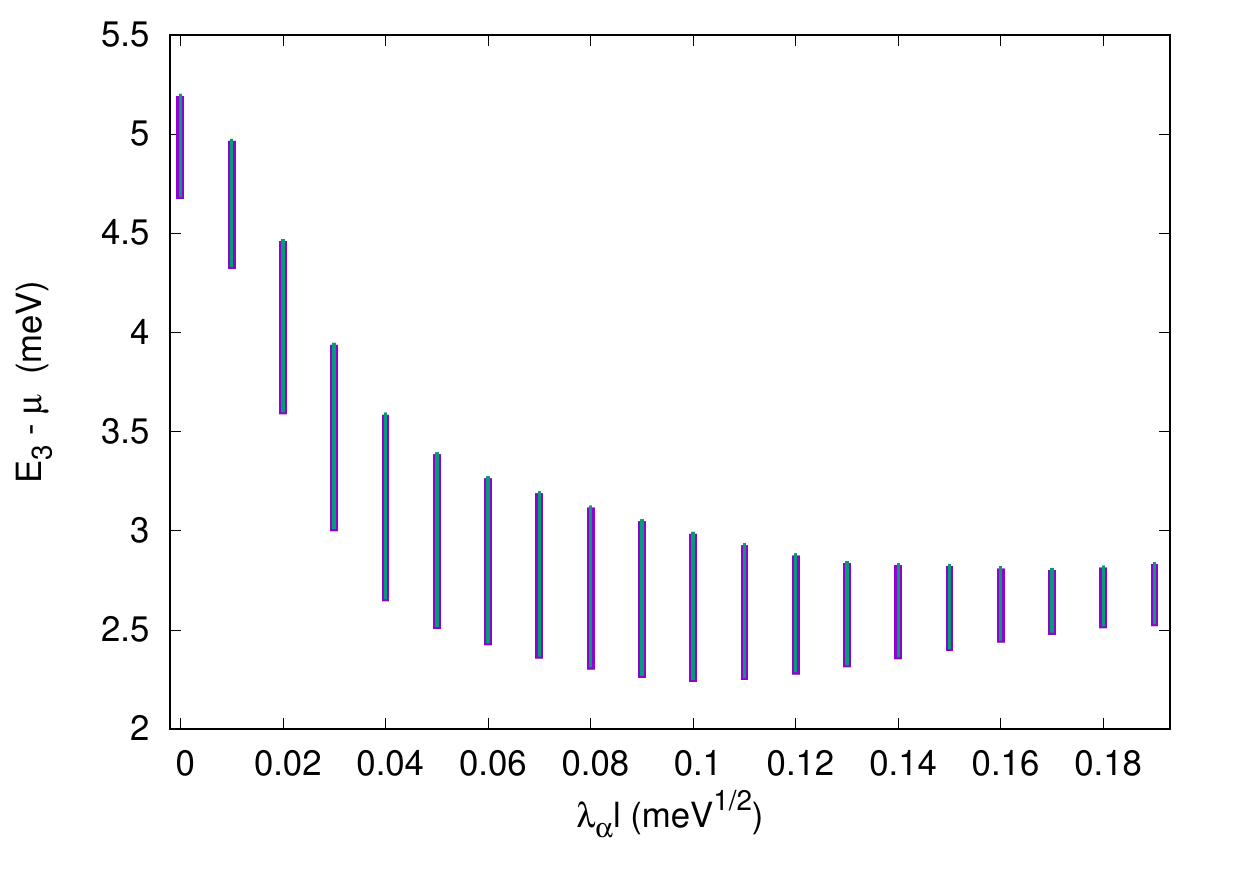}\\
	\includegraphics[width=0.48\textwidth,bb=0 00 400 300]{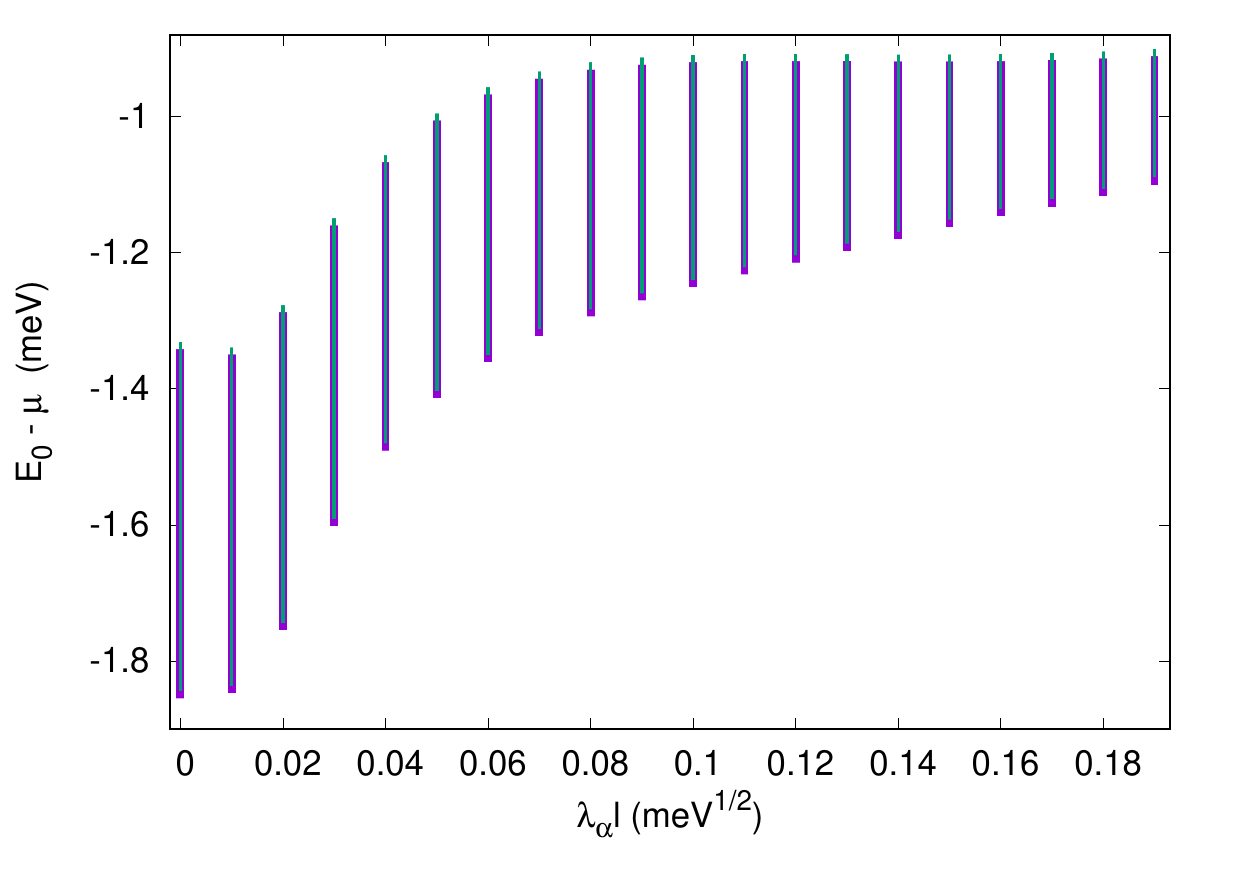}
	\includegraphics[width=0.48\textwidth,bb=0 00 400 300]{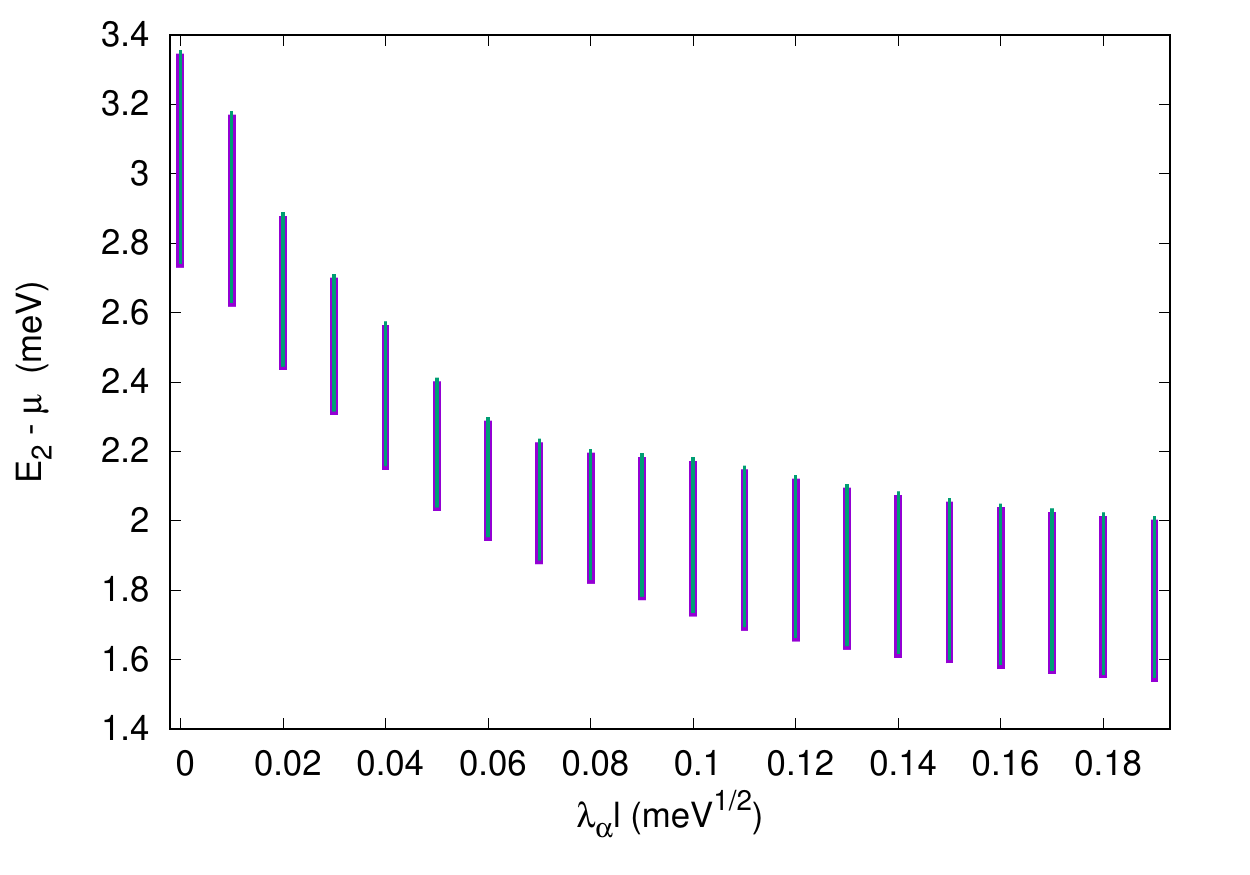}
	\caption{The dispersion of the 4 lowest energy bands as a function of the electron-photon
		     coupling. The two overlapping spin components for the lowest band (bottom left),
		     the second band (top left), the third band (bottom right), and the fourth one
		     (top right) are displayed with two different colors. $pq=1$, and $\mu$ is the
             chemical potential at $T=1.0$ K. $\hbar\omega_\alpha =1.0$ meV.}
	\label{Edisp}
\end{figure*}

In this self-consistent QEDFT model of a 2DEG interacting with the modes of a
photon cavity the chemical potential $\mu$ shifts nontrivially with the interaction
strength. This is not unexpected for DFT calculation, but in Fig.\ \ref{Edisp} we
plot $(E_{\bm{\beta\theta}\sigma}-\mu)$ for the 8 lowest energy bands (shown in Fig.\ \ref{Eband})
as a function of the electron-photon coupling strength $\lambda_\alpha l$.
We remember that for $pq=1$ the band dispersion is much larger than the Zeeman spin splitting,
so in each panel of Fig.\ \ref{Edisp} there are two bands. The lower left panel shows the
two bands with highest occupation, and the other panels show higher lying bands, that are
almost unoccupied, as here, $T=1.0$ K giving the thermal energy $k_\mathrm{B}T\approx 0.086$ meV.

In a many-body calculation of interacting electrons and photons one utilizes a
Fock-basis of states that include both states from the original electron and photon
bases \cite{ANDP:ANDP201500298,Flick2017}. The resulting states are referred to as
photon dressed states, and commonly the term photon replicas is used.
Here, the DFT orbitals or states are derived in a different way, but it may still be
appropriate to talk about photon-dressed electrons, that are fermions, but no
photon replicas are identifiable. But a look at the left panels of Fig.\ \ref{Edisp} should remind us
of another phenomena, a {\sl FIR magnetoplasmon polariton}. The reduction of the gap between
the energy bands seen in the left panels of Fig.\ \ref{Edisp} is reminiscent of the formation of
a bosonic magnetoplasmon polariton composed of a cavity photon mode and the polarization of
charge across a gap in the energy spectrum.
{We emphasize, that our DFT-results can not be taken as a proof of the emergence of a
magnetoplasmon polariton, but we take them as an indication. Additionally, we point out that concurrently to the
narrowing of the bandgap or gaps the mixing of the states of the original basis functions increases beyond what
is caused by the Coulomb interaction alone. The electron-photon interaction is leading to mixing of
energy bands as is seen in the formation of a magnetoplasmon polariton.}
Corresponding structure is found in the energy
spectrum of the 2DEG for higher values of the magnetic flux $pq$.

\begin{figure}[htb]
	\centerline{\includegraphics[width=0.36\textwidth,bb=54 50 339 300]{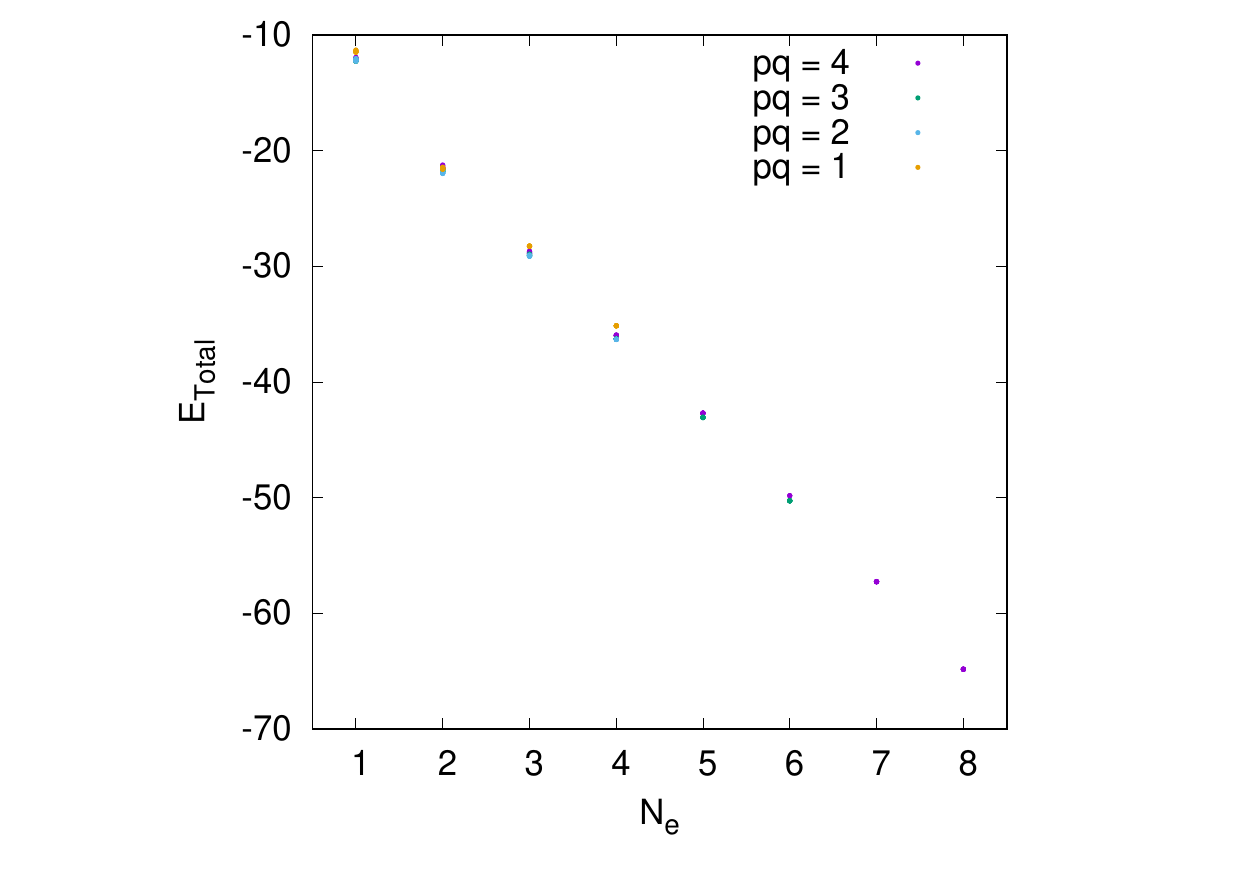}}
	\centerline{\includegraphics[width=0.44\textwidth,bb=00 50 400 300]{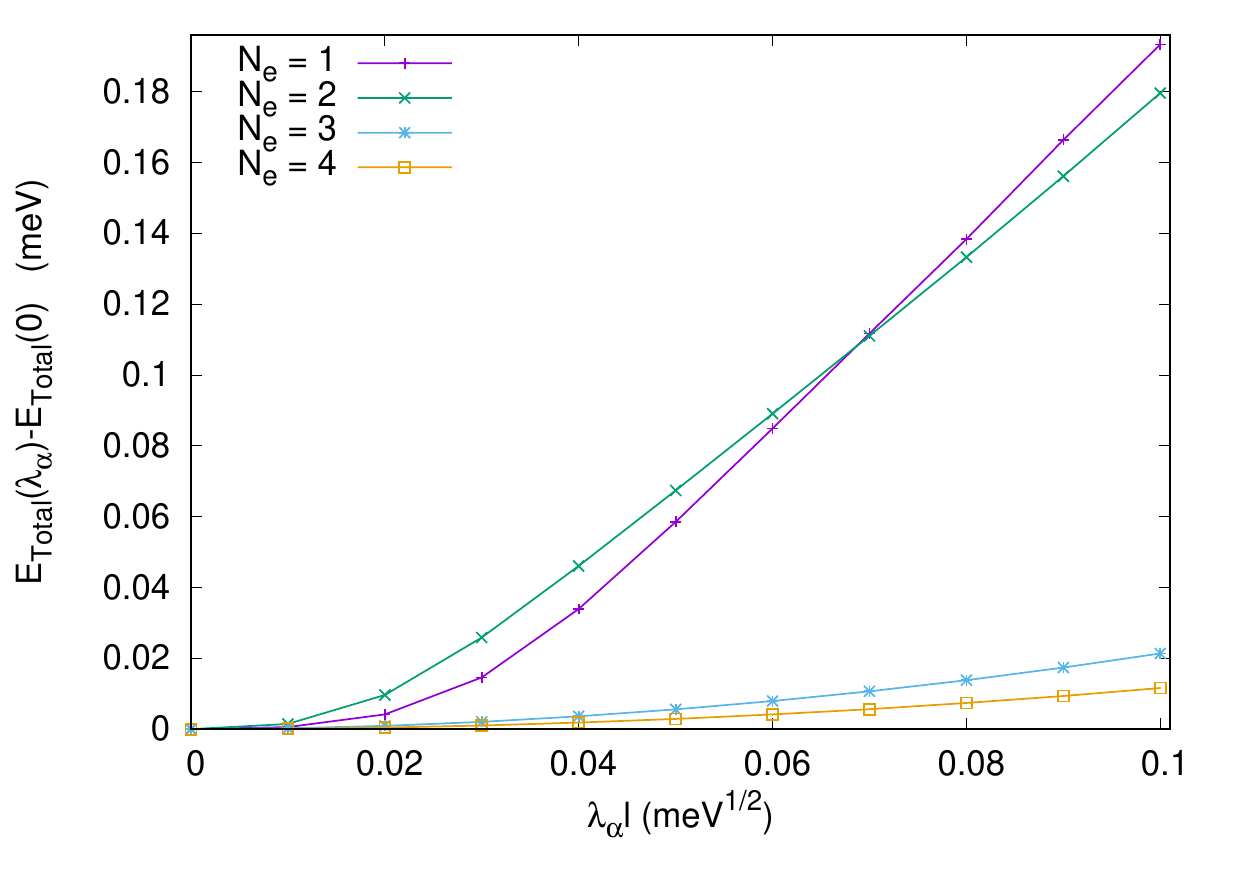}}
	\centerline{\includegraphics[width=0.44\textwidth,bb=00 00 400 300]{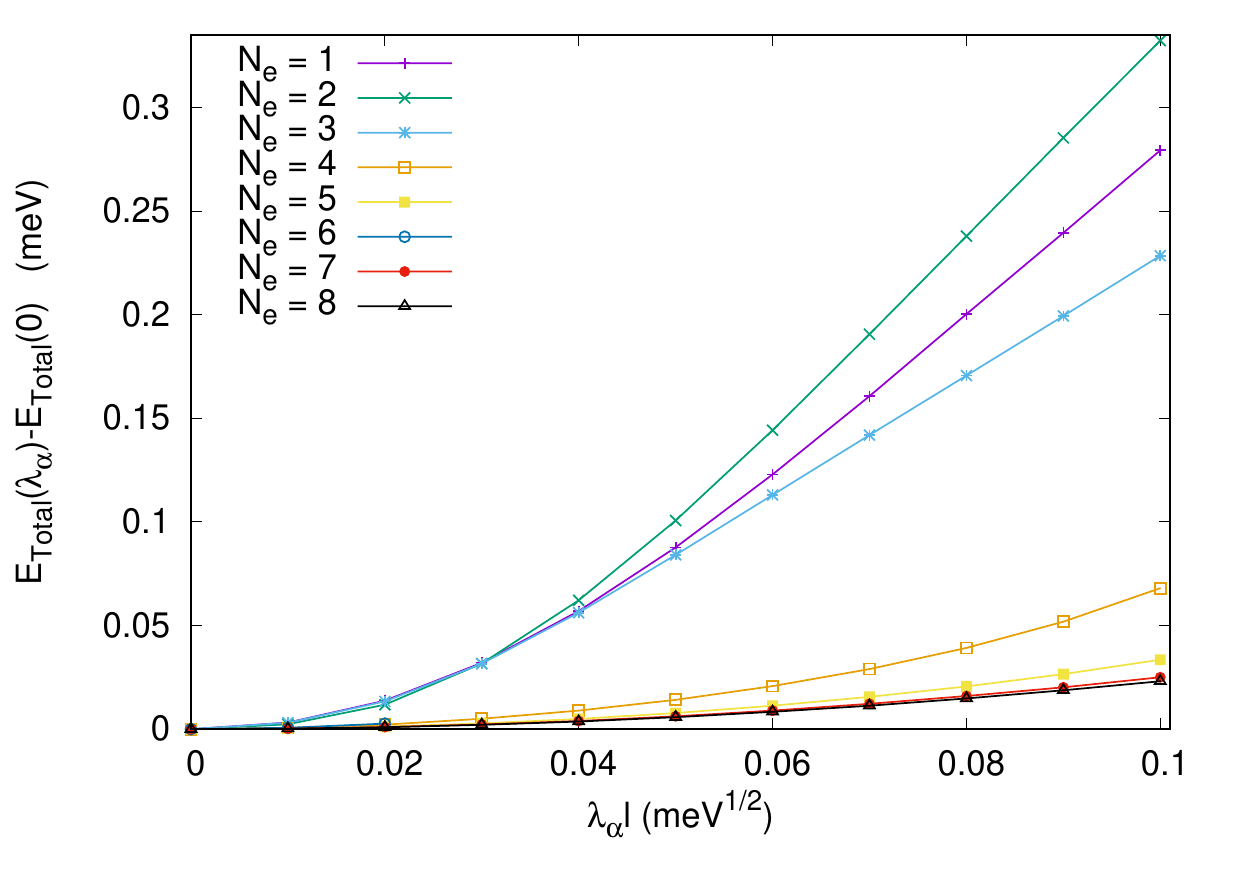}}
	\caption{The total energy $E_\mathrm{Total}$ as function of the number of electrons in each
	         unit cell $N_e$ for 4 different values of magnetic flux quanta $pq$
             and the electron-photon coupling strength $\lambda_\alpha l$ in the range from 0 - 0.1
             meV$^{1/2}$ (top panel). The difference in the total energy
             $E_\mathrm{Total}(\lambda_\alpha )-E_\mathrm{Total}(0)$
             as a function of  $\lambda_\alpha l$ for $pq=1$ (center panel), and $pq=4$ (bottom
             panel).
             The spectrum in the top panel contains points for all values of the coupling
             strength $\lambda_\alpha l$ used for the construction of the lower panels.
             $T=1.0$ K, and $\hbar\omega_\alpha =1.0$ meV.}
	\label{Et}
\end{figure}

The electron-photon coupling is expected to increase the total energy for the
electronic system, but in a dot modulated charge neutral 2DEG this can be hidden
by other details. In Fig.\ \ref{Et} we display in the top panel the total energy
as a function of the number of electrons in a unit cell $N_e$, but also for the
electron-photon coupling in the range $\lambda_\alpha l= 0 - 0.1$ meV$^{1/2}$.
Clearly, it is difficult to discern the effects of the coupling here.
They become clearer in the center panel for $pq=1$, and the bottom panel for $pq=4$,
where we display $E_\mathrm{Total}(\lambda )-E_\mathrm{Total}(0)$ as a function of  $\lambda_\alpha l$
for several values of $N_e$. As expected the electron-photon coupling leads to an increase in
the total energy of the system. The largest effects for the increase of the total energy
is for $N_e=1$ or 2 for $pq=1$ (center panel). The increase for $N_e=3$ or 4 is much smaller,
what can be referred back to the decreased polarizability of the electron charge for the higher
number of electrons. Here, plays in the availability of neighboring states to facilitate the
polarization and the ``shell structure'' of the quantum dots. For $pq=4$ (bottom panel) similar
behavior is found, but there the increase in the total energy is still rather high for $N_e=3$,
which reflects the change in the shell structure with a higher magnetic field.

\begin{figure}[htb]
	\includegraphics[width=0.48\textwidth,bb=00 10 400 300]{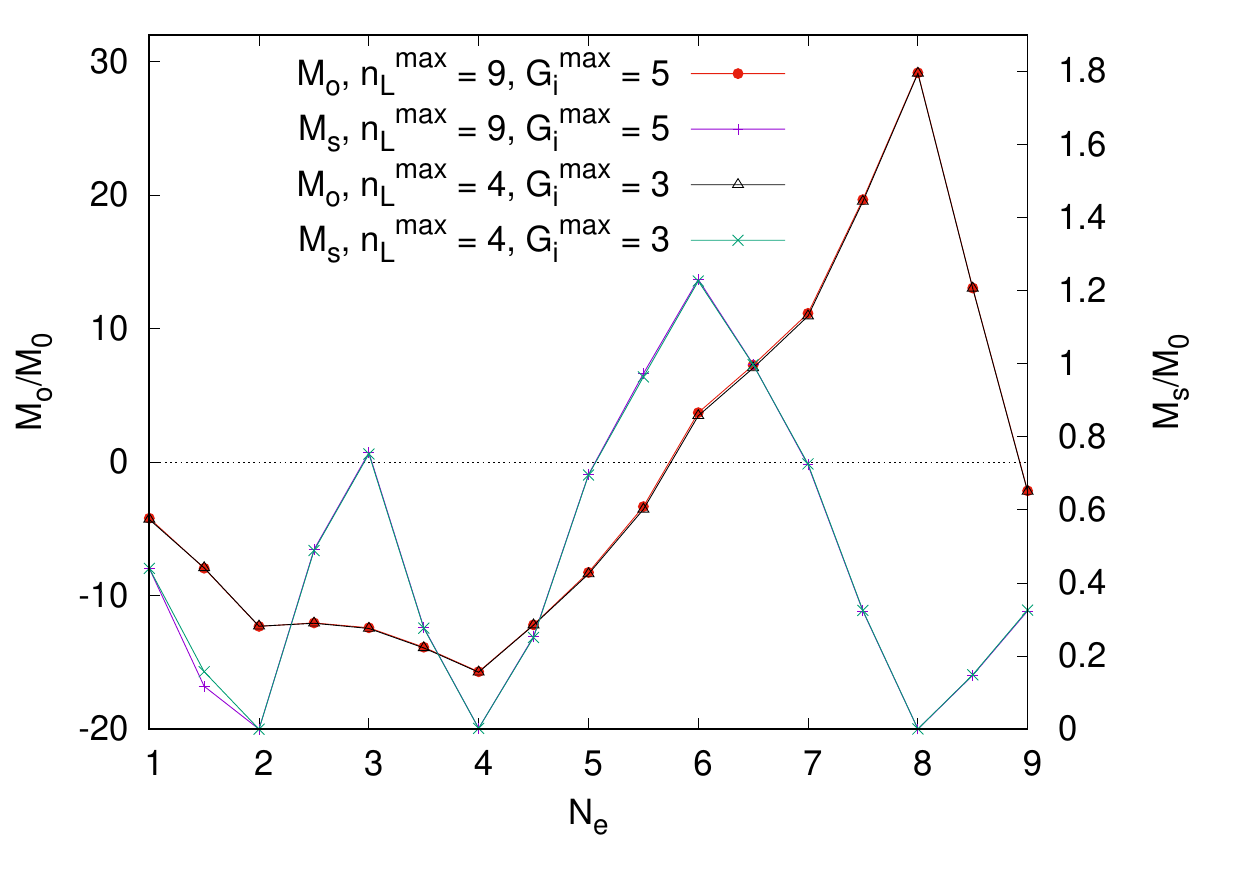}
	\caption{The orbital $M_\mathrm{o}$ and the spin component of the magnetization
			$M_\mathrm{s}$ as functions of the number of electrons in a unit cell
			$N_e$. $\lambda_\alpha l=0$, $T=1.0$ K,
		    $g^*=0.44$ and $M_0=\mu^*_\mathrm{B}/L^2$.}
	\label{Magn0}
\end{figure}

The magnetization for a dot modulated 2DEG without the coupling to cavity modes
is presented in Fig.\ \ref{Magn0}. Note the difference in scales for the orbital
component and the spin component due to the GaAs parameters used.
In Fig.\ \ref{Magn0} we show that in absence of the electron-photon interaction
relatively few Fourier coefficients are need for the calculation. The notation
$G_i^\mathrm{max}=5$ indicates that the integers $G_1$ and $G_2$ specifying the reciprocal
lattice vectors, $\bm{G} = G_1\bm{g}_1 + G_2\bm{g}_2$, are in the interval $-5,\dots,0,\dots,5$.
For the calculations with $\lambda_\alpha l\neq 0$ we use $\max |G_i|=12$ for $pq=1$, and 15 for
higher values of the flux $pq$.

The spin component of the magnetization, $M_\mathrm{s}$, in Fig.\ \ref{Magn0} for $pq=4$ vanishes
for $N_e=2$, 4, 8, corresponding to the filling factors $\nu =1/2, 1$, and 2, respectively.
For a system with no Coulomb interaction between the electrons $M_s$ also vanishes for
$N_e=6$, but here all parts of the Coulomb interaction, the direct one and the exchange and correlation
contributions, cause a rearrangement of the spin structure around $N_e=6$ corresponding to
the filling factor $\nu = 3/2$. This happens in the same region as the orbital magnetization,
$M_\mathrm{o}$ changes sign. $M_\mathrm{o}$ assumes local extrema values for the filling
factors $\nu =1$ and 2.

\begin{figure*}[htb]
	\includegraphics[width=0.48\textwidth,bb=0 50 336 350]{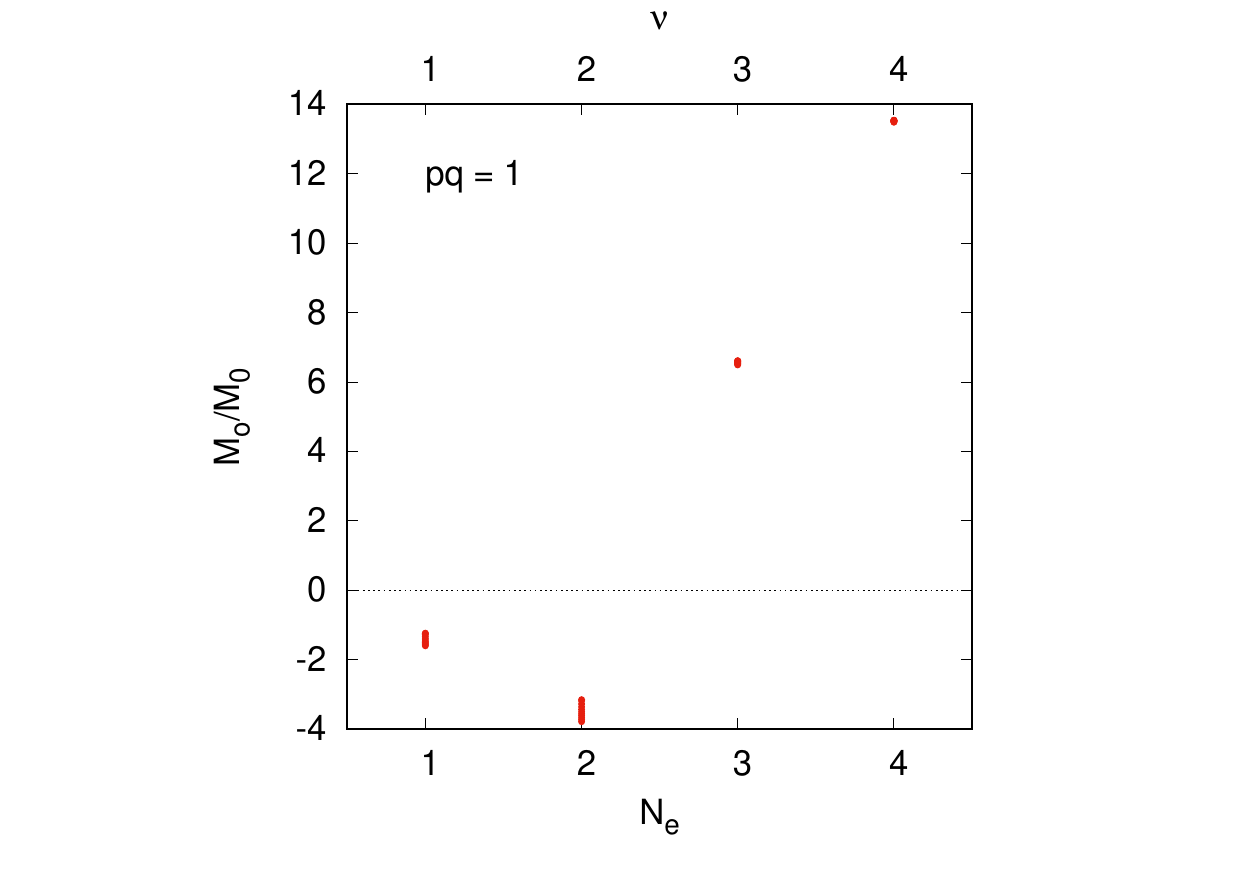}
	\includegraphics[width=0.48\textwidth,bb=0 50 336 300]{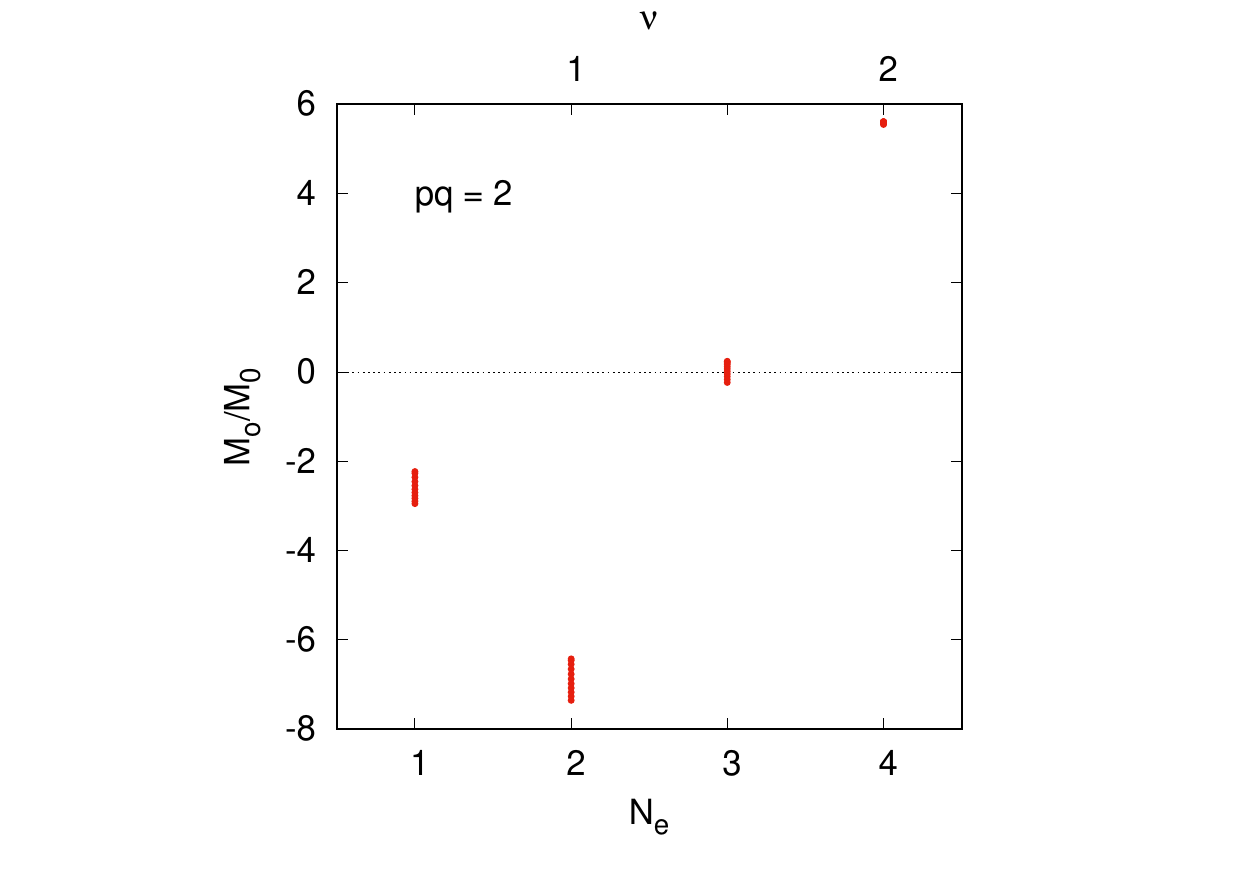}\\
	\includegraphics[width=0.48\textwidth,bb=0 00 336 300]{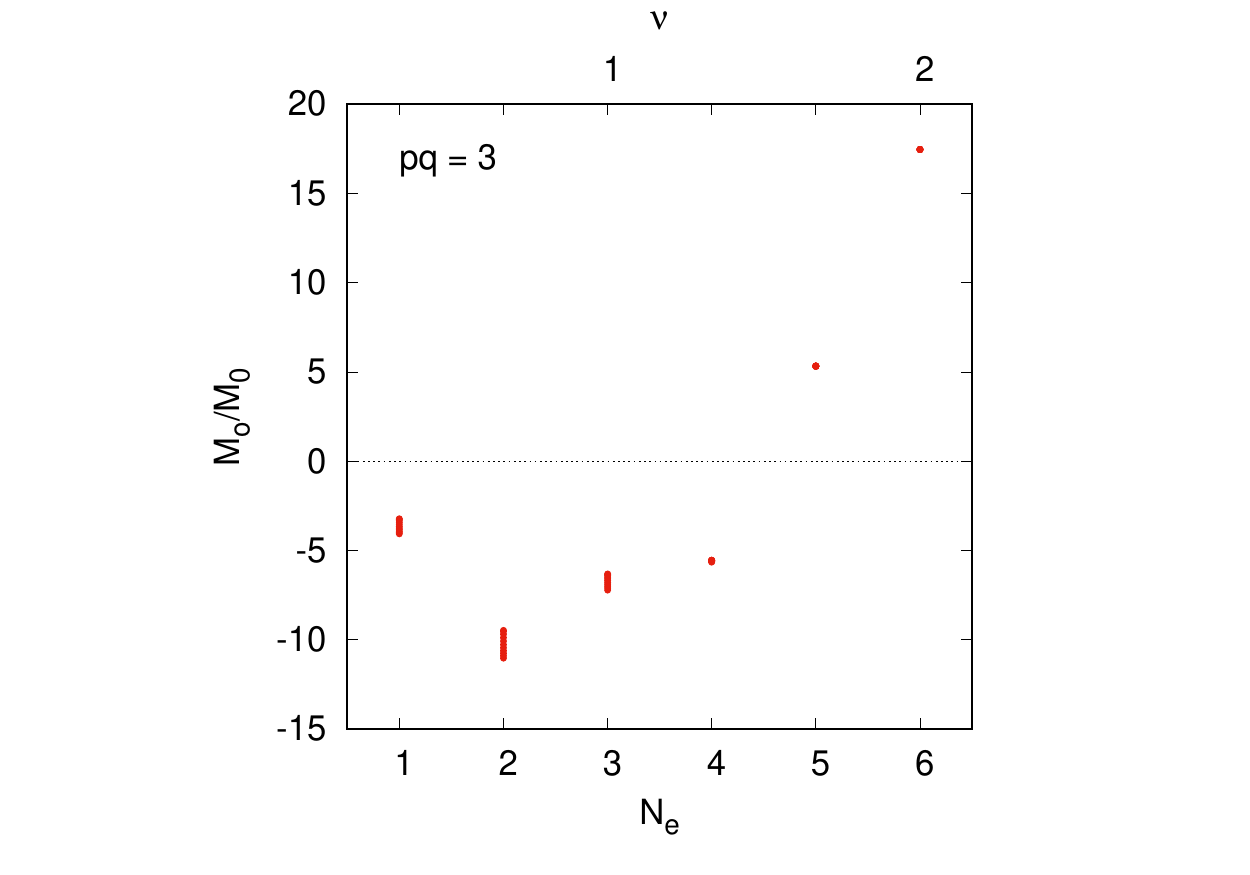}
	\includegraphics[width=0.48\textwidth,bb=0 00 336 300]{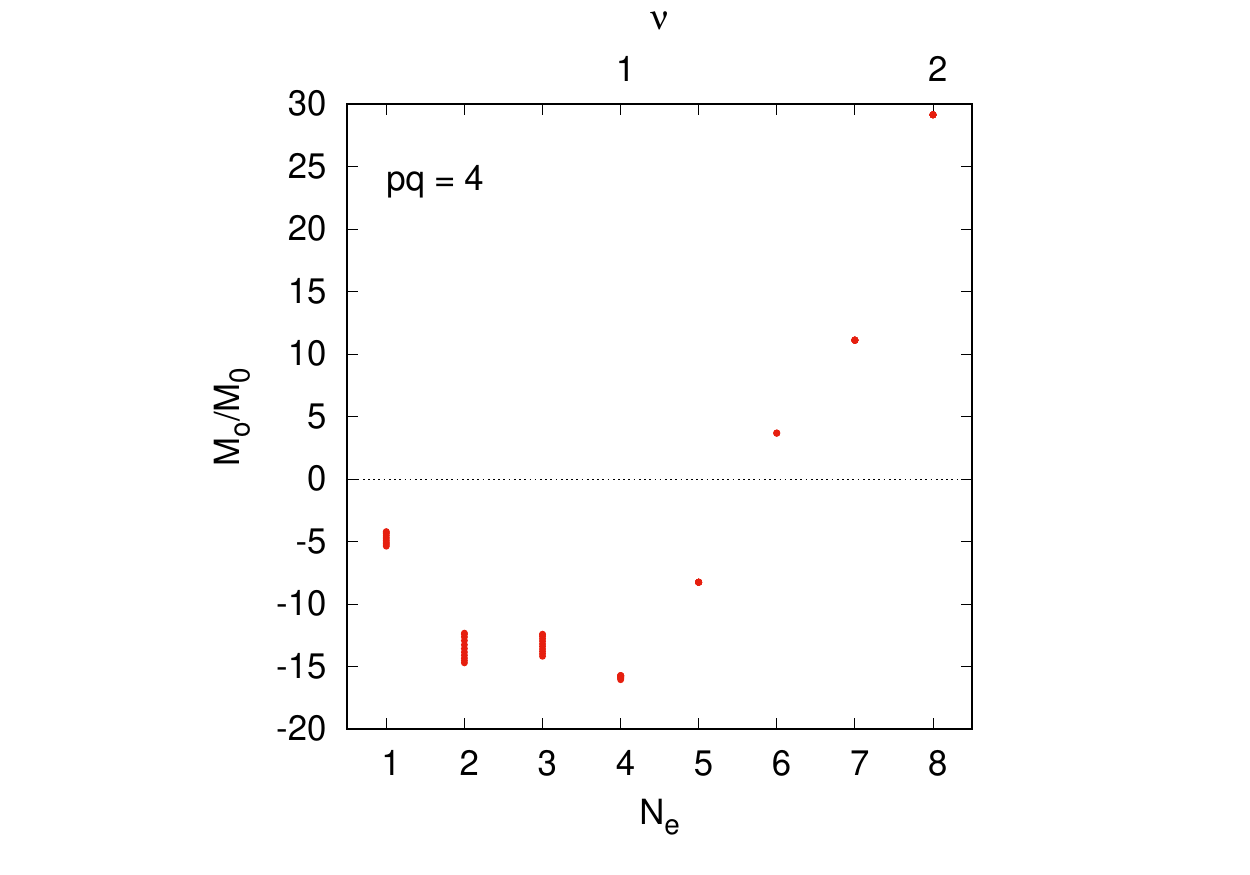}
	\caption{The orbital magnetization $M_\mathrm{o}$ as a function of the number
		     of electrons in a unit cell $N_e$ for $pq=1$ (upper left), $pq=2$
		     (upper right), $pq=3$ (bottom left), and $pq=4$ (bottom right)
		     for the electron-photon coupling strength $\lambda_\alpha l$ in the range from 0 - 0.1
		     meV$^{1/2}$. The filling factor $\nu$ is indicated by the top horizontal
		     axes. $T=1.0$ K, $\hbar\omega_\alpha =1.0$ meV,
		     and $M_0=\mu^*_\mathrm{B}/L^2$.}
	\label{MagnNe}
\end{figure*}

As the spin contribution to the magnetization is small for the GaAs parameters we shall
in what follows concentrate on the effects of the electron-photon coupling on the orbital
magnetization. In Fig.\ \ref{MagnNe} we display, for an overview, $M_\mathrm{o}$ as a function
of $N_e$ for values of the electron-photon coupling strength $\lambda_\alpha l$ in the range
of 0 - 0.1 meV$^{1/2}$ for 4 different values of the magnetic flux $pq$. The filling factor
$\nu$ is indicated on the upper abscissa of the subfigures.

\begin{figure*}[htb]
	\includegraphics[width=0.48\textwidth,bb=0 50 400 300]{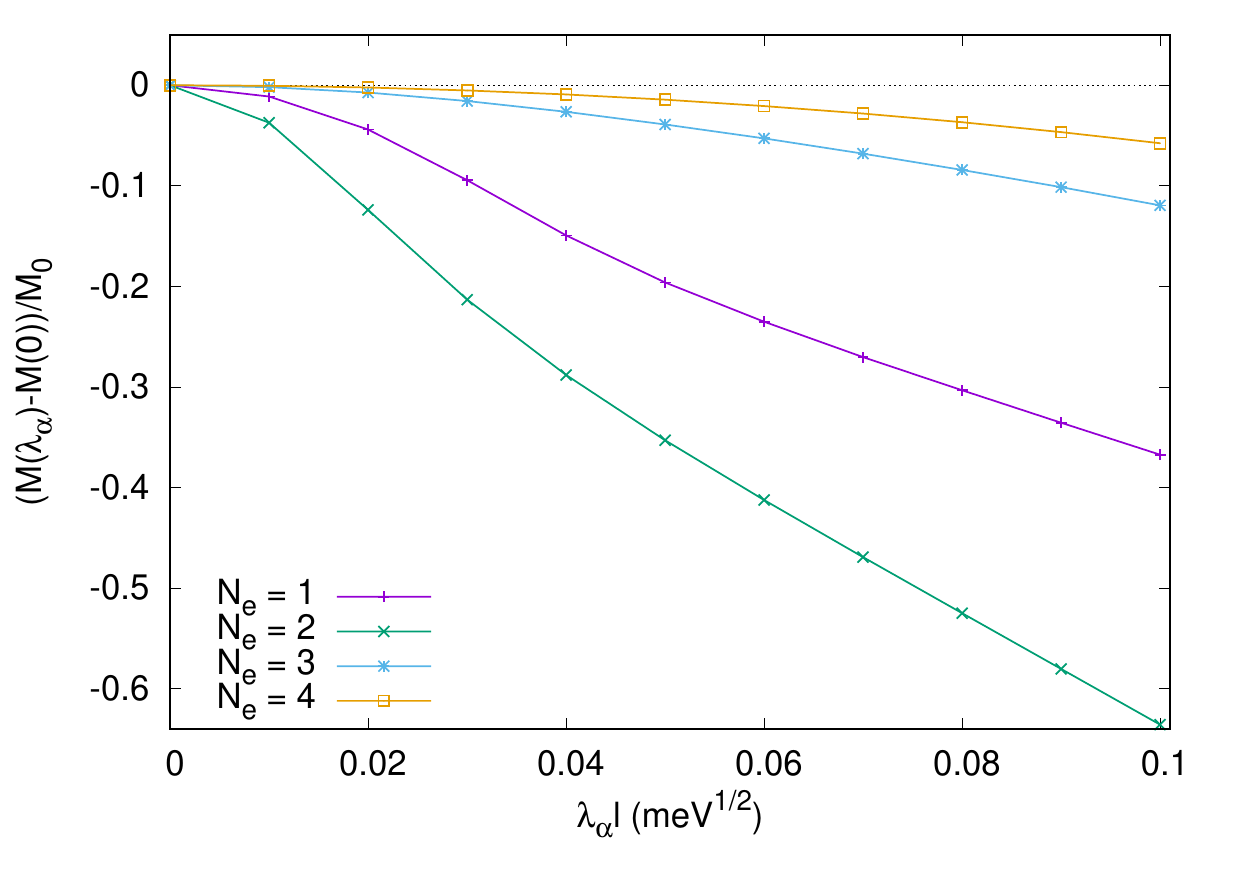}
	\includegraphics[width=0.48\textwidth,bb=0 50 400 300]{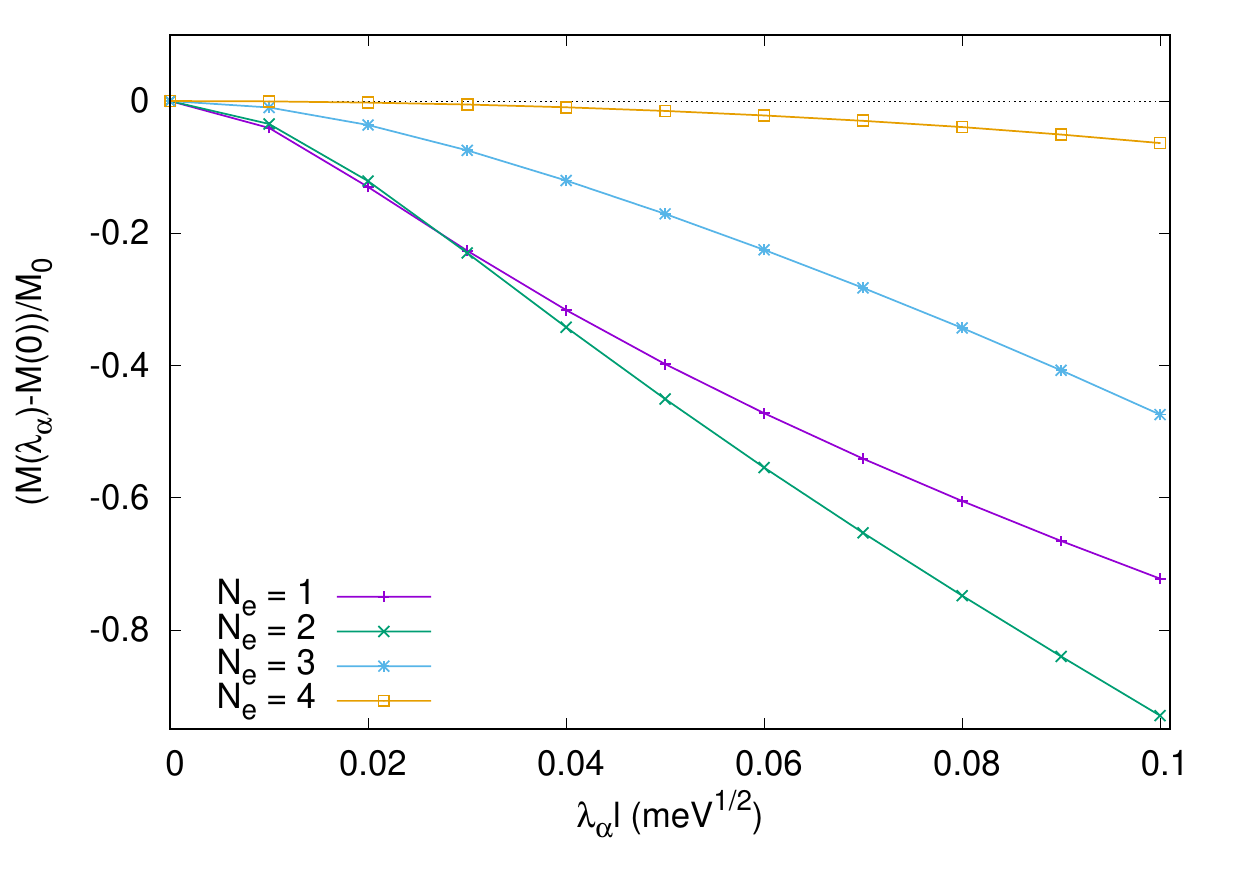}\\
	\includegraphics[width=0.48\textwidth,bb=0 00 400 300]{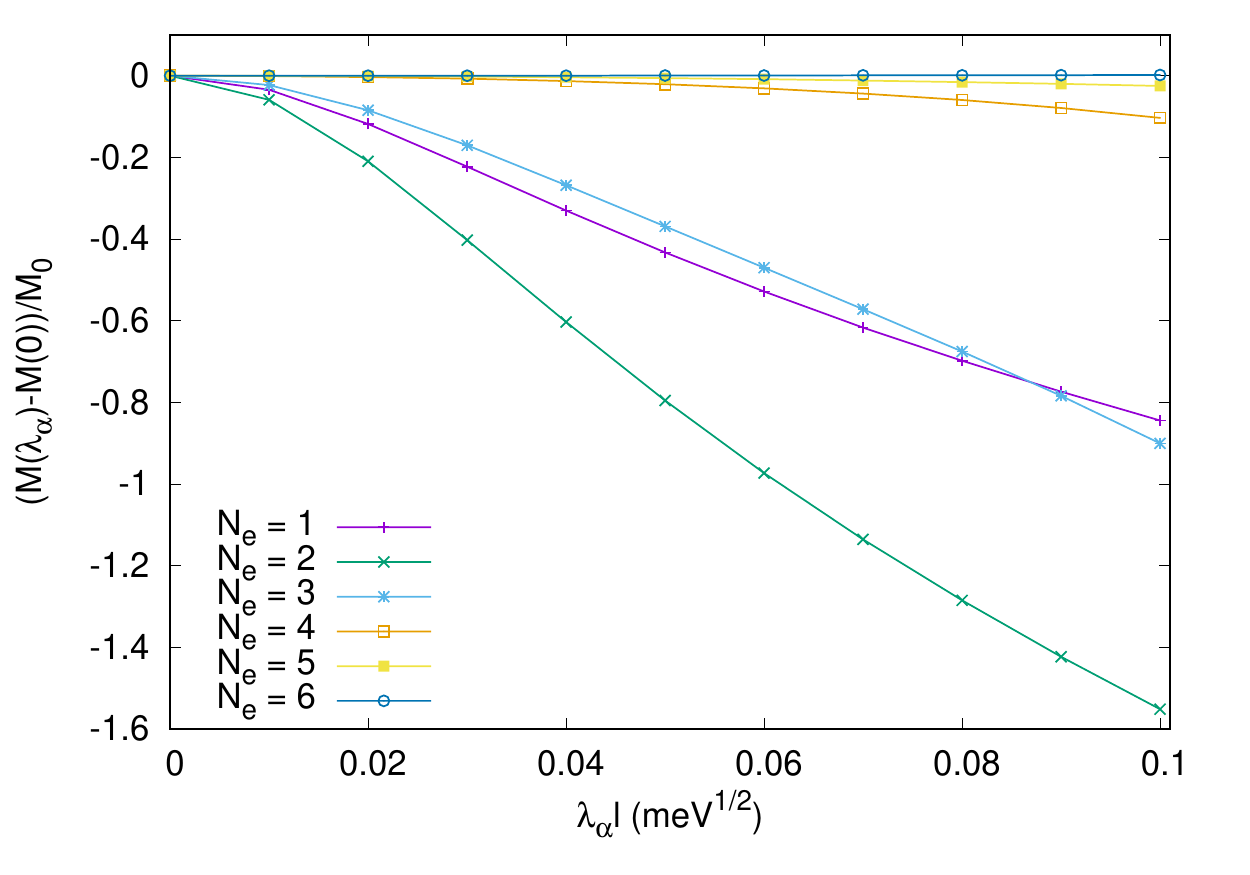}
	\includegraphics[width=0.48\textwidth,bb=0 00 400 300]{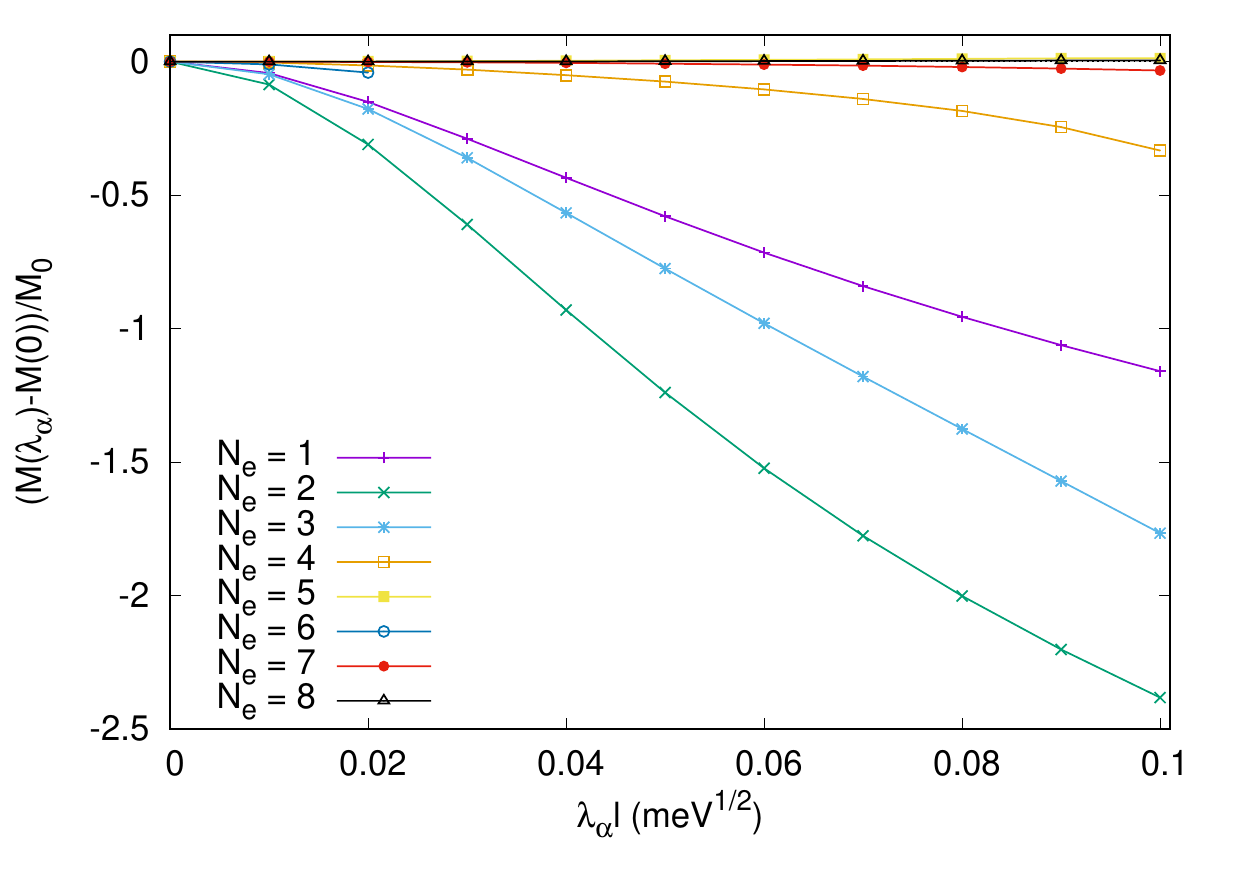}
	\caption{The difference in the orbital magnetization
		    $M_\mathrm{o}(\lambda_\alpha)-M_\mathrm{o}(0)$ as a function of the
	        electron-photon coupling strength $\lambda_\alpha l$ for $pq=1$ (upper left),
            $pq=2$ (upper right), $pq=3$ (bottom left), and $pq=4$ (bottom right).
            $T=1.0$ K, $\hbar\omega_\alpha =1.0$ meV,
            and $M_0=\mu^*_\mathrm{B}/L^2$.}
	\label{MagnLa}
\end{figure*}

Clearly, $M_\mathrm{o}$ is in certain cases influenced by the coupling $\lambda_\alpha l$
to a considerable amount, as could be expected from the structure of the integrand
for the orbital term in Eq.\ \ref{M_OS}. In order to make these changes more visible
we plot $\Delta M_\mathrm{o}=M_\mathrm{o}(\lambda_\alpha)-M_\mathrm{o}(0)$ as functions of
$\lambda_\alpha l$ for 4 different values of the magnetic flux $pq$ and the several values
of $N_e$ in Fig.\ \ref{MagnLa}.
Like for the total energy displayed in Fig.\ \ref{Et} we notice that the largest
changes in the orbital magnetization occur for 1, 2, or 3 electrons in each quantum
dot, but here the largest change is always found for $N_e=2$ with increasing coupling
$\lambda_\alpha l$. The sign of $\Delta M_\mathrm{o}(\lambda_\alpha)$ can understood with a comparison
with the information in Fig.\ \ref{MagnNe}.

 We note that the shape of the curves for
$\Delta M_\mathrm{o}(\lambda_\alpha)$ is nontrivial. It depends on the filling factor, the
magnetic flux $pq$, and the structures of the energy bands.
A noteworthy feature of the orbital magnetization for a QD array characterized by small values
of $N_e$ is the sensitivity to the increase of $\lambda_{\alpha}l$, even in the weakly interacting
regime. Indeed, for all four values of $pq$ one notices visible changes in the magnetization when
the electron-photon coupling slightly increases to $\lambda_{\alpha}l=0.02$.
One also notices that for $N_e=1,..,3$ the deviations from the values corresponding to the
non-interacting regime $\lambda_{\alpha}l=0$ are considerably enhanced as $pq$ increases.
On the other hand, the number of magnetic flux quanta $pq$ also influences the `response' of
the magnetization to the electron-photon coupling. For example, as $pq$ increases
the magnetization of the three-electron QD array shows a stronger dependence on
$\lambda_{\alpha}l$. This behavior suggests that in such a complex system the contribution
of the coupling strength to equilibrium properties depends also on the configuration of the system,
that is charge distribution, energy bands, gaps and magnetic field.

Even though only single-photon
exchange processes were taken into account with terms up to second order in $\lambda_\alpha l$
in the construction of the exchange and correlation functional for the electron-photon
interaction (\ref{spin-d-d}), the self-consistency required in the DFT-calculation forms
higher order processes built from the the single-photon exchange processes. On top of this
the underlying square dot lattice and the Coulomb interaction between the electrons have
a strong influence.

\section{Conclusions}
\label{Conclusions}
A self-consistent quantum-electrodynamical density-func\-tional approach was developed and
implemented for a square periodic quantum dot lattice placed in between the parallel plates of a far-infrared photon micro-cavity and subjected to an external perpendicular homogeneous magnetic field. The Coulomb interaction and the electron-photon coupling are treated self-consistently,
while the exchange-correlation functionals are adapted for the GaAs 2DEG hosting the array.
We explore how the interplay between the Coulomb interaction and the electron-photon coupling
affects some equilibrium properties of the system.

In the presence of the cavity the charge density of the system is polarized and its total
energy increases with increasing electron-photon coupling. The orbital magnetization of the electron system changes in nontrival ways, depending on the magnetic flux, the number of electrons in
each dot and their bandstructure. The dispersion of the energy bands closest to the chemical potential
separating empty and filled states shows {an indication for} the emerging structure
of magnetoplasmon-polaritons
in the dots. The magnetization of the system is especially adequate to explore in future
experiments as it is an equilibrium measurable quantity, only depending on its static properties.
The orbital magnetization could thus be a good quantity to test the quality of the exchange and
correlation functionals for the electron-photon interaction.

\begin{acknowledgments}
This work was financially supported by the Research Fund of the University
of Iceland, and the Icelandic Infrastructure Fund. The computations were performed on resources
provided by the Icelandic High Performance Computing Center at the University of Iceland.
V.\ Mughnetsyan and V.G.\ acknowledge support by the Armenian State Committee
of Science (grant No 21SCG-1C012). V.\ Mughnetsyan acknowledges support by the Armenian State
Committee of Science (grant No 21T-1C247).
V.\ Moldoveanu acknowledges financial support from the Romanian Core Program PN19-03
(contract No.\ 21 N/08.02.2019).
\end{acknowledgments}

%----------------------------------------------------------------------------------------
%
\appendix

\section{Electron - electron Coulomb exchange and correlation functionals}
\label{e-Coulomb}
The local spin density (LSDA) functional is a functional of the spin components
of the electron density $n_e=n_\uparrow + n_\downarrow$, and the electron spin
polarization expressed as $\zeta = (n_\uparrow - n_\downarrow)/n_e$. The functional has been
interpolated between the nonpolarized and the ferromagnetic limit by Tanatar and
Ceperley \cite{Tanatar89:5005} for 2DEG. The form of the interpolation in 2D
relies on work of Barth and Hedin \cite{Barth_1972}, Perdew and Zunger \cite{PhysRevB.23.5048}
\begin{equation}
	f^{i}(\zeta)=\frac {(1+\zeta)^{3/2}+(1-\zeta)^{3/2}-2}{2^{3/2}-2}.
\label{fi-zeta}
\end{equation}
In an external magnetic field it is more natural to replace the electron density with the local
filling factor $\nu ({\bf r}) = 2\pi l^2n_e({\bf r})$ and the spin components thereof
\cite{Lubin97:10373,Gudmundsson03:0304571}. The exchange and correlation functional is
then \cite{Koskinen97:1389}
\begin{equation}
      	\epsilon_\mathrm{xc}^B(\nu , \zeta)=\epsilon_\mathrm{xc}^{\infty}
      	(\nu)e^{-f(\nu)}+\epsilon_\mathrm{xc}^0
      	(\nu, \zeta)(1-e^{-f(\nu)}),
\end{equation}
where the interpolation between vanishing magnetic field and an infinite one depends on
$f(\nu )= (3\tilde\nu /2) + 7\nu^4$ and the high field limit is
$\epsilon_\mathrm{xc}^{\infty}(\nu)=-0.782\sqrt{\nu}e^2/(\kappa l)$. The low magnetic field limit
of the functional is
\begin{equation}
	\epsilon_\mathrm{xc}^0(\nu,\zeta)=\epsilon_\mathrm{xc}(\nu,0)+f^i(\zeta)
	\left[ \epsilon_\mathrm{xc}(\nu,1)-\epsilon_\mathrm{xc}(\nu,0)\right].
\end{equation}
The exchange and correlation parts of the functional are separated,
$\epsilon_\mathrm{xc}(\nu,\zeta)=\epsilon_\mathrm{x}(\nu, \zeta)
+ \epsilon_\mathrm{c}(\nu,\zeta)$, with $\epsilon_\mathrm{x}(\nu,0)=-[{4}/(3\pi)]
\sqrt{\nu}{e^2}/(kl) $, and
$\epsilon_\mathrm{x}(\nu,1)=-[{4}/(3\pi)] \sqrt{2\nu} e^2/(kl)$.
The parameterization of Ceperley and Tanatar for the
correlation contribution part is expressed as \cite{Tanatar89:5005}
\begin{equation}
	\epsilon_\mathrm{c}(\nu,\zeta)=a_0 \frac {1+a_1 x}{1+a_1 x+a_2 x^2+a_3 x^3}
	{Ry}^{\ast},
\end{equation}
where $x=\sqrt{r_s}=( 2 /\nu )^{1/4}( l / a_B^{\ast})^{1/2}$,
and $a_B^{\ast}$ is the effective Bohr radius. Optimized values
for the correlation parameters $a_i$ have been derived by Ceperley and Tanatar
from a Monte Carlo calculation for the 2DEG \cite{Tanatar89:5005}.
In the new variables, $\nu$ and $\zeta$, the exchange and correlation potentials
are conveniently expressed as \cite{Lubin97:10373}
\begin{align}
	V_{\mathrm{xc},\uparrow}&=\frac {\partial}{\partial \nu}
	(\nu \epsilon_\mathrm{xc})+(1-\zeta)\frac{\partial}
	{\partial \zeta}\epsilon_\mathrm{xc}\nonumber\\
	V_{\mathrm{xc},\downarrow}&=\frac {\partial}{\partial \nu}
	(\nu \epsilon_\mathrm{xc})-(1+\zeta)\frac {\partial}
	{\partial \zeta}\epsilon_\mathrm{xc}.
\label{Vxc-Coul}
\end{align}

\section{The exchange and correlation functionals for the electron-photon interaction}
\label{e-EM-functionals}
The electron-photon interaction with one photon mode labeled with $\alpha$ is described by
\begin{align}
	H_\mathrm{e-EM} = -&\frac{1}{c}\int d\bm{r}\; \bm{j}(\bm{r})\cdot\bm{A}_\alpha (\bm{r})\nonumber\\
	                  +& \frac{e^2}{2m^*c^2}\int d\bm{r}\; n_e(\bm{r})\bm{A}_\alpha^2 (\bm{r})
\label{JA-nA2}
\end{align}
with the current density
\begin{align}
	\bm{j}_{i\sigma}(\bm{r}) = -\frac{e}{m^*(2\pi)^2}\sum_\alpha\int_{-\pi}^{\pi} d\bm{\theta}\;
	\Re&\left\{ \psi_{\bm{\beta\theta}\sigma}^*(\bm{r})\bm{\pi}_i \psi_{\bm{\beta\theta}\sigma}(\bm{r}) \right\}\nonumber\\ &f(E_{\bm{\beta\theta}\sigma}-\mu),
\label{currD}
\end{align}
where $i = x,y$ indicates its component along the cartesian spatial directions.
We consider a parallel plates microcavity with the 2DEG plane in its center.
As the wavelength of the FIR cavity photons is much larger than the lattice
length $L$ we assume the the cavity vector potential in the dipole approximation
to be described by
\begin{equation}
	\bm{A}_\alpha = \bm{e}_\alpha{\cal A}_\alpha\left(a^\dagger_\alpha + a_\alpha \right),
\label{Ag-field}
\end{equation}
where $\bm{e}_\alpha$ is the polarization vector of the photon mode $\alpha$ with
energy $\hbar\omega_a$. We will here assume the polarization uniform, as might be
accomplished by a cylindrical cavity or in a rectangular cavity with equal horizontal
lengths. Within these approximations the electron-photon interaction can thus be
expressed as
\begin{align}
	  H_\mathrm{e-EM} = \hbar\omega_c\left[l\bm{I}\cdot\bm{e}_\alpha
	  \left\{\left(\frac{e{\cal A}_\alpha}{c}\right)\frac{l}{\hbar}\right\}
	  \left(a^\dagger_\alpha + a_\alpha \right)\right.  \nonumber\\
      + N_e \left\{\left(\frac{e{\cal A}_\alpha}{c}\right)\frac{l}{\hbar}\right\}^2
      \left. \left(a^\dagger_\alpha + a_\alpha \right)^2   \right]
\label{He-EM}
\end{align}
with the dimensionless spatial integral over the current density
\begin{align}
	  l\bm{I}_i =& \frac{1}{(2\pi)^2}\sum_{\bm{\beta}\sigma}\int d\bm{r}\:\int_{-\pi}^\pi d\bm{\theta}\nonumber\\
	  &\Re \left\{ \psi_{\bm{\beta\theta}\sigma}^*(\bm{r})\left(\frac{l\bm{\pi}_i}{\hbar}\right)
	  \psi_{\bm{\beta\theta}\sigma}(\bm{r}) \right\}
	  f(E_{\bm{\beta\theta}\sigma}-\mu),
\end{align}
where $i=x,y$.
The ratio $e/m^*$  characterizing the electromagnetic coupling is hidden in the cyclotron energy
$\hbar\omega_c$, and the dimensionless quantity $(e{\cal A}_\alpha/c)(l/\hbar)$
appears in Eq.\ (\ref{He-EM}) for both the para- and the diamagnetic terms.

In the QEDFT literature the electron-photon interaction within the dipole approximation
is usually written as \cite{doi:10.1073/pnas.1518224112,flick2021simple} for each
photon mode
\begin{equation}
	H_\mathrm{int} = -\omega_\alpha q_\alpha \left(\bm{\lambda}_\alpha\cdot\bm{R} \right)
	+ \frac{1}{2}\left(\bm{\lambda}_\alpha\cdot\bm{R} \right)^2,
\label{QEDFT-Hint}
\end{equation}
{with}
\begin{equation}
	q_\alpha = \sqrt{\frac{\hbar}{2\omega_\alpha}}\left(a^\dagger_\alpha + a_\alpha \right).
\label{q_a}
\end{equation}
The coupling constant of for the photon mode $\alpha$ for the electrons and the
photons $\bm{\lambda}_\alpha$ includes the polarization of the
vector potential and has the dimension of a square root of energy over length.
A more common approach to the dipole approximation would have led us to
replace $\bm{I}$ with the mean value of the spatial coordinate $\bm{R}$
\begin{equation}
	l^2\bm{I} \rightarrow \bm{R}.
\end{equation}
together with the identification
\begin{equation}
      \left|\bm{\lambda}_\alpha\right| = \frac{\hbar\omega_c}{\sqrt{\hbar\omega_\alpha}}
      \frac{\sqrt{2}}{l} \left\{\left(\frac{e{\cal A}_\alpha}{c}\right)\frac{l}{\hbar}\right\}
\end{equation}
in order to make a connection to our presentation of the electron-photon interaction
(\ref{He-EM}).
Here, we will thus determine the coupling $\lambda_\alpha l$ in units of $\sqrt{\mathrm{meV}}$.

For the derivation of the exchange-correlation functional including single-photon exchange
processes and terms up to $\sim\lambda_\alpha^2$ we point out the discussion of Flick
\cite{flick2021simple}, but after inspection of the dynamic electron polarizability
$\alpha_{\mu\nu}(i\omega)$ introduced in Eq.\ (8) of Ref.\ \cite{flick2021simple} we select
a dynamic polarizability for the isotropic 2DEG as
\begin{equation}
	\alpha (iu) = \frac{\hbar}{4\pi}\int d\bm{r}\; \frac{\hbar\omega_p}{\left(\omega_p^2/3 + \omega_g^2 + u^2\right)\hbar^2}
\label{aiu}
\end{equation}
in order to obtain the needed dimensionality. The frequency integration for the exchange and
correlation functionals expressed by Eq.\ (7) and (9) in Ref.\ \cite{flick2021simple} then
leads to the functional $E_\mathrm{xc}^\mathrm{GA}[n,\bm{\nabla}n]$ displayed in Eq.\ (\ref{E_EM_xc}).

{
One, may question the use of the dipole approximation for the electron-photon interaction
for an extended 2DEG in an external magnetic field. The feature lengths of the 2DEG system, the
lattice length $L$ of the quantum array and the cyclotron radius $l$ are much smaller than the
wavelength of the photons in the FIR-cavity, which is in the range of several tenths of
micrometers. Secondly, one has to have in mind how the FIR-absorption
single dots or wires \cite{Gudmundsson95:17744}, or arrays of quantum dots
\cite{Gudmundsson96:5223R,Dahl90:5763,Dahl93:15480} was successfully calculated earlier using
the dipole interaction and compared to experiments.
There, the external exciting field had no spatial variation radially,
but an angular pattern exciting dipole or higher order electrical modes, but
the self-consistent local field correction acquired spatial variations due to the underlying
system, that resulted in, for example, the excitation of Bernstein and higher order modes.
}

%----------------------------------------------------------------------------------------
%
%\balance
\frenchspacing
%\bibliographystyle{apsrev4-2}
%\bibliography{mod_qd}

%
%
%----------------------------------------------------------------------------------------
%
\end{document}